\documentclass[12pt ]{article}
\usepackage{amsmath}
\usepackage{graphicx}
\usepackage{hyperref}
\usepackage{lmodern}
\usepackage{amssymb}
\usepackage{booktabs}
\usepackage{tikz}
\usetikzlibrary{patterns}
\usepackage{pdfpages}
\usepackage{dsfont}
\usepackage{bbm}

\newcommand{\be}{\begin{equation}}
\newcommand{\ee}{\end{equation}}

\newcommand{\beqa}{\begin{eqnarray}}
\newcommand{\eeqa}{\end{eqnarray}}
\newcommand{\nn}{\nonumber}

\def\CD {{\cal D}}
\def\CE {{\cal E}}
\def\CF {{\cal F}}
\def\CG {{\cal G}}
\def\CH {{\cal H}}
\def\CI {{\cal I}}

\def\CL {{\cal L}}

\def\CN {{\cal N}}
\def\CO {{\cal O}}

\begin{document}

\newpage

\setlength{\baselineskip}{7mm}
\begin{titlepage}
 
\begin{flushright} 
 {\tt NRCPS-HE-73-2025} 
\end{flushright}

\begin{center}
{\Large ~\\{\it   QCD Effective Lagrangian  \\ 
and \\
Condensation of Chromomagnetic Flux Tubes}
}

\vspace{3cm}

{\sl George Savvidy

\centerline{${}$ \sl Institute of Nuclear and Particle Physics}
\centerline{${}$ \sl Demokritos National Research Center, Ag. Paraskevi,  Athens, Greece}

}
 
\end{center}
\vspace{3cm}

\centerline{{\bf Abstract}}

We compute the effective  action for covariantly constant gauge fields that are solutions of the sourceless Yang-Mills equation and have the form of  magnetic flux tubes. They represent a superposition of infinite many alternating monopole/ani-monopole pairs situated at infinity,  with each pair having a structure similar to  the Nielsen-Olesen magnetic flux tube.   The  chromomagnetic flux tubes condensation is stable and indicates that the Yang-Mills vacuum state is highly degenerate.    
  \vspace{12pt}

\noindent

\end{titlepage}

\pagestyle{plain}

\tableofcontents

\section{\it Introduction}

In this article we compute the effective  action for covariantly constant gauge fields that are solutions of the sourceless Yang-Mills equation and have the form of  magnetic flux tubes. The covariantly constant gauge fields describe a superposition of infinite many alternating monopole/ani-monopole pairs situated at infinity,  with each pair having a structure similar to  the Nielsen-Olesen magnetic flux tube  \cite{Nielsen:1973cs,Nambu:1974zg} but without presence of any Higgs field.  
Importantly,  the effective action is a gauge-invariant functional  for  sourceless gauge fields and  has a universal form  similar  to  the Lagrangian for the constant gauge field. The   Yang-Mills vacuum state is highly degenerate with the vacuum field configurations ranging from a constant gauge field  to  a rich chromomagnetic flux tube structure permeating the space in all directions.

 The covariantly constant gauge fields are solutions of the equation
\be\label{YMeqcovint1}
\nabla^{ab}_{\rho} G^{b}_{\mu\nu} =0
\ee
 and are also the solutions of sourceless Yang-Mills equation.  The well known solution of the equation (\ref{YMeqcovint1}) is the constant Abelian  field  : 
\be\label{consfield1}
A^{a}_{\mu} = - {1\over 2} F_{\mu\nu} x_{\nu}   n^a, 
\ee
where $F_{\mu\nu} $ and $n^a$, $ n^{a} n^{a} =1$ are space-time constants.   The general solutions of the equation (\ref{YMeqcovint1}) were found recently and are obtained through the nontrivial space-time dependence of the unit colour vector $n^a(x)$ within the following  ansatz  \cite{ tHooft:1974kcl, Corrigan:1975zxj, Cho:1979nv, Biran:1987ae, Faddeev:1998eq, Faddeev:2001dda, Savvidy:2024sv, Savvidy:2024ppd, Savvidy:2024xbe}:
\be\label{choansatzint}
A^{a}_{\mu} =  B_{\mu} n^{a}  +
{1\over g} \varepsilon^{abc} n^{b} \partial_{\mu}n^{c},
\ee
where $B_{\mu}(x)= A^a_{\mu} n^a$  is an Abelian gauge field and  
$
n^{a} n^{a} =1,
$
$n^{a} \partial_{\mu} n^{a} =0.
$
The field-strength tensor for the gauge fields (\ref{choansatzint}) factorises:
\be\label{chofactint}
G^{a}_{\mu\nu} (A)=   G_{\mu\nu}~ n^a(x), ~~~~~~~~~G_{\mu\nu}=  F_{\mu\nu} + {1\over g} S_{\mu\nu},
\ee
where
$$
F_{\mu\nu}= \partial_{\mu} B_{\nu} - \partial_{\nu} B_{\mu}, ~~~~~~~~~~
S_{\mu\nu}= \varepsilon^{abc} n^{a} \partial_{\mu} n^{b} \partial_{\nu} n^{c}.
$$
The general solution of the equation (\ref{YMeqcovint1}) in terms of unit vector $n^a$ is \cite{Savvidy:2024sv, Savvidy:2024ppd, Savvidy:2024xbe}\footnote{Considering the ansatz with a magnetic charge $g_m$ of the following form $A^{a}_{\mu} =  B_{\mu} n^{a}  +
g_m \varepsilon^{abc} n^{b} \partial_{\mu}n^{c},$ one can get convinced that it is solution of the Yang-Mills equation only when $g_m = 1/g$.}:
\be\label{generasolint}
n^a(\vec{x})= \{\sin \theta(X)  \cos\Big({Y \over \theta(X)^{'} \sin \theta(X)} \Big),~\sin \theta(X)  \sin\Big({Y \over \theta(X)^{'}  \sin \theta(X) }\Big),~ \cos \theta(X)   \}, 
\ee 
where $X= a_{\mu}  x_{\mu} \equiv (a\cdot x) $,  $Y=(b\cdot x)$ and $a_{\mu}, b_{\nu}$ are a constant four-vectors. The explicit form of the vector potential $A^a_{\mu}$ is obtained by substituting the  unit colour vector (\ref{generasolint})  into (\ref{choansatzint}).  These are the exact solutions of the Yang-Mills equation in the background field $F_{\mu\nu}(B)$ and have the non-Abelian term $S_{\mu\nu}(n)$ induced by the unit vector $n^a$.  

The tensor structure of the solution (\ref{choansatzint}) is similar to the spherically symmetric  point-like Wu-Yang solution $n^a = \frac{x^a}{r}$ \cite{Wu-Yang, Wu:1975es, Wu:1967vp, Wu:1976if, Wu:1975vq}  while here the gauge field  is homogeneously distributed all over the 3d-space.  The physical meaning of the solution is that it describes a superposition of infinite many alternating monopole/ani-monopole pairs situated at infinity,  with each pair having a structure similar to  the Nielsen-Olesen magnetic flux tube that covers the whole 3d-space. 

The moduli space of the solutions (\ref{generasolint}), (\ref{choansatzint}) is defined by the $\theta(X)$ function. In, particular, when $\theta(X)=\arcsin({1\over  \cosh(a \cdot x)})$, we obtain the "hyperbolic" solution:
\be\label{hypersol12}
n^a(x)= \{ {\cos((b\cdot x) \cosh^2(a\cdot x)) \over  \cosh(a\cdot x)}, { \sin((b\cdot x) \cosh^2(a\cdot x)) \over \cosh(a\cdot x)}, \tanh(a\cdot x)  \},
\ee 
when $\theta(X)= a_{\mu} x_{\mu} \equiv (a\cdot x) $, we will obtain the  "trigonometric" solution:
\be\label{ansatz3}
n^a(\vec{x})= \{ \sin a x \cos \Big({b y \over \sin a x }\Big),~\sin a x \sin\Big({b y \over \sin a x}\Big),~ \cos a x  \}.
\ee 
and finally, considering  $\theta(X)=\arcsin(\sqrt{1- (a\cdot x)^2})$ we obtain the "polynomial"  solution:
\be\label{polsol12}
n^a(x)= \{ \sqrt{1-(a\cdot x)^2} \cos(b\cdot x),~ \sqrt{1-(a\cdot x)^2} \sin(b\cdot x),~ (a\cdot x)  \}
\ee 
representing a magnetic flux wall of a finite thickness $  2/\vert a \vert $. All these solutions have a constant energy density:
\be 
 \epsilon =  {1\over 4 }G^{a}_{ij} G^{a}_{ij}  =   {(g  \vec{H} -   \vec{a} \times \vec{b} )^2 \over 2 g^2},
\ee
where $B_i = - \frac{1}{2} F_{ij}x_j$, $a_{\mu}=(0,\vec{a})$, $b_{\nu}=(0,\vec{b})$. The chromomagnetic flux $\Phi$ defined as \cite{tHooft:1981bkw, tHooft:1979rtg,tHooft:1980xss} 
\be\label{magfluxint}
A(L)={1\over 2} Tr P \exp{(i g \oint_L  \hat{A}_k d x^k)}  = \cos(\frac{1}{2} g \Phi)
\ee
is equal to $\Phi_1={2\pi \over g}$ and  $A(L_1) =-1$ when a closed contour $L_1$ is surrounding  a cell of an oriented magnetic flux tube of the square area $  { 2\pi \over a b}$ in the $(x,y)$ plane of the polynomial solution (\ref{polsol12}).   The flux through the contour $L_2$  of a nearby  cell of the same area $ { 2 \pi \over a b}$ is negative.  The chromomagnetic fluxes have  {\it opposite orientations} in the nearby cells. This fact is illustrated by computing the total flux through the  union of two  cells $L_1\cup L_2$,  which vanishes $\Phi(L_1\cup L_2) =0$ and $A(L_1\cup L_2)=1$ (see Figures \ref{fig1}, \ref{fig11} and Appendix B for details).

We  compute the effective Lagrangian for the chromomagnetic flux tube solution (\ref{choansatzint}), (\ref{generasolint}). We found that the effective Lagrangian  on chromomagnetic flux tube configurations has a universal form   and  is a Lorentz- and gauge-invariant functional depending on  two invariants, $\CF = {1\over 4}G^a_{\mu\nu}G^a_{\mu\nu}=   {\vec{\CH}^2_a - \vec{\CE}^2_a \over 2}$ and  $\CG = {1\over 4}G^a_{\mu\nu}G^{*a}_{\mu\nu} =  \vec{\CE}_a \vec{\CH}_a $.  We conclude  that the Yang-Mills vacuum state is highly degenerate with the condensate of chromomagnetic flux tubes.

The article is organised as follows. In the second section we will present  a general solution of the covariantly constant field equation  (\ref{YMeqcovint1}) and will analyse its properties in the third   section.  The properties of the conserved current $J^a_{\mu} = g \epsilon^{abc} A^b_{\nu} G^c_{\nu\mu}$   and of the corresponding current vorticity $\omega^a_i = \epsilon_{ijk} \partial_j J^a_k $  supporting the solution geometry will be analysed.  The geometry  has a lattice cell structure of the alternating chromomagnetic flux tubes periodically repeating themselves with oppositely orientated fluxes in the  neighbouring  cells  (see Fig.\ref{fig1} and Fig.\ref{fig11}).

In the forth section we will review the basic properties of the effective Lagrangian in the Yang-Mills theory and will prove the gauge invariance of the effective action for sourceless gauge fields.  {\it The importance of having exact solutions of the sourceless Yang-Mills equation lies in the fact that only in that case the vacuum polarisation and the effective Lagrangian represent the gauge-invariant physical effects}  \cite{Savvidy:1977, Batalin:1976uv, Batalin:1979jh}.

The computation  of the effective Lagrangian can be reduced to the  evaluation of the  matrix elements of the operator  $U(s) = \exp\{-i H s\}$  \cite{Schwinger:1951nm}.   
The matrix elements of the operators  $U(s)$  can be computed by three alternative methods \cite{Savvidy:1977, Batalin:1976uv}. In the first method suggested by Schwinger in QED one can consider the  operator $ H $   as the Hamiltonian of a  "particle" moving in a background field with the "particle"  space-time  coordinates $x_{\mu}(s)$  depending on the proper time $s$ \cite{Schwinger:1951nm} and the equation of motion in the Heisenberg representation.  In the second method the matrix elements are computed by using the path-integral representation   \cite{Batalin:1970it, Savvidy:1977}, and in the third method the determinant is computed as a product of the eigenvalues, as in the original article of Heisenberg and Euler \cite{Heisenberg:1936nmg, tHooft:1976snw}.  These methods will be used in this article. 

In the  fifth section we reexamine the properties of the effective Lagrangian for the constant gauge field (\ref{YMeqcovint1}) stressing that  the effective Lagrangian  is a Lorentz- and gauge-invariant functional depending on  two invariants, $\CF $ and  $\CG $.   

In the sixth section we discuss the presence/absence of imaginary terms in the effective Lagrangian.  The significance of the presence/absence of imaginary terms in the effective Lagrangian is connected with the fact that they define the quantum-mechanical stability of the  field configurations \cite{Nielsen:1978rm}. A number of physical arguments and analytical results  leads to the conclusion that there are no  imaginary terms in the effective action  for chromomagnetic fields \cite{Savvidy:2022jcr, Savvidy:2023kft, Savvidy:2023kmx}. The underlying physical reason lies in the fact that the magnetic field does no work  and therefore cannot separate  a pair of virtual vacuum charged particles  to the asymptotic states at infinity \cite{Savvidy:1977}, as it happens in the case of the electric field \cite{Sauter:1931zz, Heisenberg:1936nmg, Schwinger:1951nm}.  The vacuum persistence probability  \cite{Schwinger:1951nm}  should be less than 1,  therefore any imaginary term in the effective action should be nonnegative  \cite{Nielsen:1978rm}:
$
|\langle 0\vert 0 \rangle|^2= \left| \exp \left\{ i \Gamma(H) \right\} \right|^{2} = \exp \left\{ -2~ \mathcal{I}m ~\Gamma(H) \right\},~~2~ \mathcal{I}m ~\Gamma(H) \geq 0. 
$
The appearance of a negative mode is a result of the quadratic approximation for quantum fluctuations and the inclusion of the quartic  self-interaction of a negative mode eliminates the instability and the imaginary term from effective action \cite{Nielsen:1978nk, Ambjorn:1978ff,Savvidy:2022jcr, Savvidy:2023kft, Savvidy:2023kmx}.   

In the seventh, eighth and ninth sections we evaluate the effective Lagrangian for flux tube solutions. The spectrum of the Hamiltonian  $H$ can be evaluated exactly, allowing to obtain  a one-loop effective Lagrangian and to demonstrate its universal form. One can conjecture that the effective Lagrangian for general chromomagnetic flux tube solutions  has this universal form. 

\section{\it  Covariantly constant   gauge fields}

The covariantly constant gauge fields were defined by the equation (\ref{YMeqcovint1}) \cite{Batalin:1976uv, Savvidy:1977as, Matinyan:1976mp, Brown:1975bc, Duff:1975ue}   and, as it was mentioned in the Introduction,
the effective action is gauge invariant on  sourceless gauge fields \cite{Savvidy:1977, Batalin:1976uv, Batalin:1979jh}.  Here we will consider the $SU(2)$ algebra; the consideration can be extended to other algebras as well. By taking the covariant derivative $\nabla^{ca}_{\lambda}$ of the l.h.s (\ref{YMeqcovint1}) and interchanging the derivatives one can obtain that
$
 [G_{\lambda\rho},  G_{\mu\nu}] =0,
$
which means that the field-strength tensor factorises into the product  of Lorentz tensor and the colour unit vector in the direction of the Cartan's sub-algebra:
\be\label{covconfac}
G^{a}_{\mu\nu}(x)= G_{\mu\nu}(x) n^a(x). 
\ee
 Both fields can depend on the space-time coordinates. A well known solution  has the following form \cite{Batalin:1976uv, Savvidy:1977as, Matinyan:1976mp, Brown:1975bc, Duff:1975ue}: 
 \be\label{consfield}
B^{a}_{\mu} = - {1\over 2} F_{\mu\nu} x_{\nu}   n^a , 
\ee
where $F_{\mu\nu} $ and $n^a$ are space-time constants and  $ n^{a} n^{a} =1$. It is convenient to call this solution "constant Abelian  field" \footnote{The  solution has six parameters $F_{\mu\nu}$, four translations $x_{\nu} \rightarrow x_{\nu} + x_{0 \nu} $ and two parameters $n^a$ in the case of the $SU(2)$ group.}.  The general solutions of the equation (\ref{YMeqcovint1})   \cite{Savvidy:2024sv, Savvidy:2024ppd,Savvidy:2024xbe} can be obtained through the nontrivial space-time dependence of the unit vector $n^a(x)$ (\ref{choansatzint}).  The field-strength tensor $G_{\mu\nu}(x)$ (\ref{chofactint}) is identical with the 't Hooft form of the electromagnetic field-strength tensor of a magnetic monopole in the Yang-Mills-Higgs model \cite{tHooft:1974kcl,Corrigan:1975zxj}:
\be\label{abelean}
G_{\mu\nu}=   n^a G^a_{\mu\nu}
+ {1\over g }\epsilon^{abc} n^a   \nabla_{\mu}  n^b \nabla_{\nu} n^c  \equiv \partial_{\mu} B_{\nu} - \partial_{\nu} B_{\mu} + {1\over g }\epsilon^{abc} n^a   \partial_{\mu}  n^b  \partial_{\nu}  n^c,~~~~~ n^a = { \phi^a \over \vert \phi \vert},
\ee 
where $\nabla_{\mu}  n^a= \partial_{\mu} n^a -g \epsilon^{abc} A^b_{\mu} n^c$,  $B_{\mu}= A^{a}_{\mu} n^a$ and the unit colour vector $n^a$ is associated with the adjoint scalar (\ref{abelean}).    The definition (\ref{abelean}) satisfies the Maxwell equations, except for the space-time point, where the scalar field vanishes, $\phi_a(x) = 0$, and  the field $n^a(x)$ develops a singularity. 
  
The covariantly constant field-strength tensor (\ref{covconfac})  has a factorisation form similar  to the one in the Yang-Mills-Higgs model (\ref{abelean}).  Here the role of the unit colour vector field $n^a(x)$ is not connected with any adjoint scalar field but with the Yang-Mills field itself instead. It is therefore natural to search the covariantly constant gauge fields  in the form (\ref{choansatzint}).   In that case (\ref{YMeqcovint1}) reduces to the following equation \cite{Savvidy:2024sv, Savvidy:2024ppd,Savvidy:2024xbe}:
 \be
   \partial_{\rho}  ( F_{\mu\nu} + {1\over g} S_{\mu\nu})=0,
 \ee
 meaning that the sum of the terms in the brackets should be a constant tensor.
It is useful to parametrise the unit vector in terms of the spherical angles:
\be\label{unitvector}
n^a = (\sin\theta \cos\phi, \sin\theta \sin\phi, \cos\theta ),
\ee
and express $S_{\mu\nu}$ in terms of spherical angles as well:
$$
S_{\mu\nu} = \sin\theta ( \partial_{\mu}  \theta  \partial_{\nu} \phi  - \partial_{\nu}  \theta   \partial_{\mu} \phi).
$$ 
Let us first consider  the solutions that have constant space components $S_{ij}$  and  $F_{ij}$ with time components  $S_{0i}$ and $F_{0i}$ equal to zero. These solutions represent a pure chromomagnetic field, and the equation (\ref{YMeqcovint1}) reduces to the following system of partial differential equations\footnote{The details concerning the solution of the equations (\ref{gen})-(\ref{ansatz7}) are given in Appendix A.}:
\beqa\label{gen}
S_{12}= \sin \theta (\partial_1 \theta \partial_2 \phi - \partial_2 \theta \partial_1 \phi ), \nn\\
S_{23}= \sin \theta (\partial_2 \theta \partial_3 \phi - \partial_3 \theta \partial_2 \phi ), \nn\\
S_{13}= \sin \theta (\partial_1 \theta \partial_3 \phi - \partial_3 \theta \partial_1 \phi ).
\eeqa
 The linear combination of these equations defines the  angle $\phi$ as an arbitrary function of  the variable 
$Y=  b_1 x +b_2 y + b_3 z -b_0 t$,
thus
$
 \phi(Y)  =\phi( b\cdot x ),
$
where $b_{\mu}, \mu=0,1,2,3$ are arbitrary real numbers. After substituting the above function into the equations (\ref{gen}) one can observe that the angle variable $\theta$ is a function of the alternative variable $X= a\cdot x $, thus
$
 \theta(X)=  \theta( a\cdot x ),
$
where $a_{\mu}, \mu=0,1,2,3$ are arbitrary real numbers as well. It follows that  the equations (\ref{gen})  reduce to the following system of differential  equations:
\be\label{genecovfie}
S_{\mu\nu} = a_{\mu}\wedge b_{\nu}  \sin\theta(X) ~ \theta(X)^{'}_X   ~ \phi(Y)^{'}_Y , 
\ee
where the derivatives are over the respective arguments.  The solutions with a constant tensor $S_{ij}$  should fulfil the following equation:
\be\label{ansatz7}
\sin\theta(X) ~ \theta(X)^{'}_X   ~ \phi(Y)^{'}_Y =1,
\ee
so that 
$
S_{\mu\nu} = a_{\mu}\wedge b_{\nu} 
$
and the square of the field-strength tensor (\ref{chofactint}) is 
 \be\label{actioncont}
 {1\over 4 }G^{a}_{\mu\nu} G^{a}_{\mu\nu} = {1\over 4 } F_{\mu\nu} F_{\mu\nu} + {a_{\mu} F_{\mu\nu} b_{\nu} \over g}  +  { a^2 b^2 - (a\cdot b)^2  \over 2 g^2}.
 \ee  
The variables in (\ref{genecovfie})  are independent, therefore we can choose the arbitrary function  $\theta$ and define the function $\phi$ by integration. Let $\theta(X) $ be an arbitrary function of $X$, then $\phi = Y/\sin \theta(X) \theta(X)^{'}_X$, and we have the following general solution for the unit vector (\ref{unitvector}):
\be\label{generasol}
n^a(\vec{x})= \{\sin \theta(X)  \cos\Big({Y \over \theta(X)^{'} \sin \theta(X)} \Big),~\sin \theta(X)  \sin\Big({Y \over \theta(X)^{'}  \sin \theta(X) }\Big),~ \cos \theta(X)   \}.
\ee 
The explicit form of the vector potential $A^a_{\mu}$ can be obtained by substituting the unit vector (\ref{generasol}) into (\ref{choansatzint}).  The arbitrary function $\theta(X)$ in the equation (\ref{generasol}) defines the moduli space of the solutions.  The singularities are located on the planes $X_s$, where  $\sin\theta(X)$  or $\theta(X_s)^{'}$ vanishes\footnote{It seems that this solution with singular surfaces can be associated  with the  singular surfaces  considered by 't Hooft in \cite{tHooft:1981bkw}, where he discussed  a possible existence of such non-perturbative solutions (see also \cite{Kapustin:2005py, Kapustin:2006pk, Gukov:2006jk, Gukov:2008sn}).}: 
\be\label{singplanes} 
\theta(X_s) = 2 \pi N, ~~~N=0,\pm 1, \pm 2...., ~~~\text{or}~~~\theta(X_s)^{'} = 0,
\ee
and  the functions $\cos\Big({Y \over \theta(X)^{'} \sin\theta(X)}\Big)$ and $\sin\Big({Y \over \theta(X)^{'} \sin\theta(X)}\Big)$ in (\ref{generasol}) are fast oscillating trigonometric functions in the vicinity of these planes while the energy density is a regular function. 
 
 The solution   (\ref{choansatzint}), (\ref{generasol}) for the vector potential $A^a_{\mu}$    depends on two variables,  $X$ and $Y$. There are two physically interesting solutions: the time-independent solutions when $a_0 =b_0= 0$ and therefore  describing stationary magnetic fluxes distributed in the 3d-space and the time-dependent solutions when $a_0 \neq 0$,  $b_0 \neq 0$ describing the propagation of chromomagnetic "strings" or "branes" when  the time components  $S_{0i}$ and $F_{0i}$ are taken to be nonzero. 
 
For the sake of transparency and compactness of the subsequent formulas we will identify this plane as the   $(x,y)$ plane by taking the vectors  $a_{\mu}=(0,a,0,0)$ and $b_{\nu}=(0,0,b,0)$, so that  $\theta(x) = f(a x)$, $\phi(x,y) = b y /f^{'}(ax) \sin f(ax) $.  The gauge field (\ref{choansatzint})   with the Abelian field $B_{1}= H y$ will take the following form:
\beqa\label{magneticsheetsolution1}
A^{a}_{i}(x,y) =   {1\over g} \left\{
\begin{array}{ccccc}   
a \Big( b y ( ({g H \over a b}-1)  \sin f    +{1 \over \sin f  } )     \cos ({b y  \over f^{'}  \sin f }) - f^{'}  \sin ({b y  \over f^{'} \sin f }) +  b y  {  f^{''}   \over f^{'2}} \cos f  \cos ({b y  \over f^{'}  \sin f }), \\ 
~~ b y ( ({g H \over a b}-1)  \sin f    +{1 \over \sin f  } )      \sin({b y  \over f^{'}  \sin f }) + f^{'}   \cos({b y  \over f^{'}  \sin f }) + b y {  f^{''}   \over f^{' 2}} \cos f  \sin ({b y  \over f^{'}  \sin f}), ~~~\\
    b  y (({g H \over a b}-1) \cos f -   { f^{''}  \over f^{'2}}  \sin f  )\Big)  \\ 
{b \over f^{'}} \Big(-\cos f  \cos ({b y  \over f^{'} \sin f}),-\cos f  \sin ({b y  \over f^{'} \sin f}), ~  \sin f \Big)\\
(0,0,0),
\end{array} \right.   
\eeqa 
where $i=1,2,3$ and the derivatives are over the whole argument $ax$. Here  $A^a_0=0$ and the singularities are at (\ref{singplanes}). One can verify explicitly  that it is a solution of the Yang-Mills equation \cite{Savvidy:2024sv, Savvidy:2024ppd,Savvidy:2024xbe}.   

When   $a_{\mu}=(0,\vec{a})$, ~$b_{\nu}=(0,\vec{b})$, the magnetic energy density has the following form (\ref{actioncont}):
\beqa\label{trigenergydens}
\epsilon(\gamma) &=&  {1\over 2 g^2} (g \vec{H} -\vec{a}\times \vec{b})^2 = {1\over 2 g^2} \Big( \vert g \vec{H} \vert^2  - 2 \vert g \vec{H}\vert \vert \vec{a}\times \vec{b}\vert  \cos\gamma  + \vert \vec{a}\times \vec{b} \vert^2 \Big)
\eeqa  
and depends on the modular parameter  $\gamma$. The minima of $\epsilon$ are realised when $ \gamma  = 0$ or $2\pi$ and the maximum at $ \gamma  = \pi$:
\be
\epsilon_{min} = {1\over 2 g^2} \Big( \vert g \vec{H} \vert  -  \vert \vec{a}\times \vec{b} \vert \Big)^2,~~~~~
\epsilon_{max} = {1\over 2 g^2} \Big( \vert g \vec{H} \vert  +  \vert \vec{a}\times \vec{b} \vert \Big)^2.
\ee
 The two minima are separated by a finite potential barrier. The zero energy density $\epsilon_{min}=0$ is realised when 
\be\label{vacuumcond}
g  \vec{H}_{vac} =  \vec{a}\times \vec{b},
\ee
and the gauge field reduces to a flat  connection $\vec{A}_{vac}=-{i \over g}  U^{-1} \vec{\nabla} U$. This takes place when three vectors  $ (\vec{H}, \vec{a}, \vec{b})$ are forming an orthogonal right-oriented frame.    At the minimum (\ref{vacuumcond}) the field-strength tensor vanishes, $G_{ij}=0$, and  the general solution   (\ref{magneticsheetsolution1}) reduces to a  flat  connection of the following form: 
\beqa\label{flatconnection} 
A^{a}_{i} &=&   {1\over g} \left\{
\begin{array}{cccc}   
 \Big( \frac{a b y}{\sin f }\cos \left(\frac{b y \csc f}{f'}\right) - a f' \sin \left(\frac{b y \csc f }{f'}\right)+ \frac{a b y f''  }{f'^2}  \cos f \cos \left(\frac{b y \csc f}{f'}\right) ,  \\ 
 \frac{a b y}{\sin f }\sin \left(\frac{b y \csc f}{f'}\right) + a f' \cos \left(\frac{b y \csc f }{f'}\right)+ \frac{a b y f'' }{f'^2}  \cos f \sin \left(\frac{b y \csc f}{f'}\right), 
  -a b y  \frac{f''  \sin f }{f'^2} \Big)  ~~~~ \\
{b\over f'}\Big(-  \cos f  \cos \left(\frac{b y \csc f }{f' }\right),- \cos f \sin \left(\frac{b  y \csc f }{f' }\right),   \sin f  \Big)&\\
(0,0,0)~,~~&
\end{array} \right. 
\eeqa 
where $a$ and $b$ are the parameters of the moduli space.  The flat connection (\ref{flatconnection}) can be represented in the standard form:
\be\label{flatconnection1}
\vec{A}_{vac} ~=~ -{i \over g}  U^{-1}(x,y)  \vec{\nabla} U(x,y).
\ee
This vacuum configuration is similar to the  CP violating  topological effect that appears due to the presence in the vacuum   field configurations that have non-vanishing  Chern-Pontryagin  index   \cite{tHooft:1976rip, Jackiw:1976pf, Callan:1976je, Jackiw:1979ur}:  
\be\label{flatcon} 
\vec{A}_n(\vec{x}) =-{i \over g} U^{-1}_n(\vec{x}) \nabla U_n(\vec{x}), ~~~~ U_1(\vec{x})= {\vec{x}^2 -\lambda^2  - 2 i \lambda \vec{\sigma} \vec{x} \over \vec{x}^2 +\lambda^2}, ~~~~~U_n = U^n_1.
\ee
The values of the gauge field  (\ref{flatcon}), although gauge equivalent to $\vec{A}(x) = 0$, are not removed from the integration over the field configurations by the gauge-fixing procedure because they belong to different topological classes and are separated by potential barriers \cite{tHooft:1976rip, Jackiw:1976pf, Callan:1976je, Jackiw:1979ur}.

\section{\it Examples of covariantly constant gauge fields}

\begin{figure}
 \centering
\includegraphics[angle=0,width=8cm]{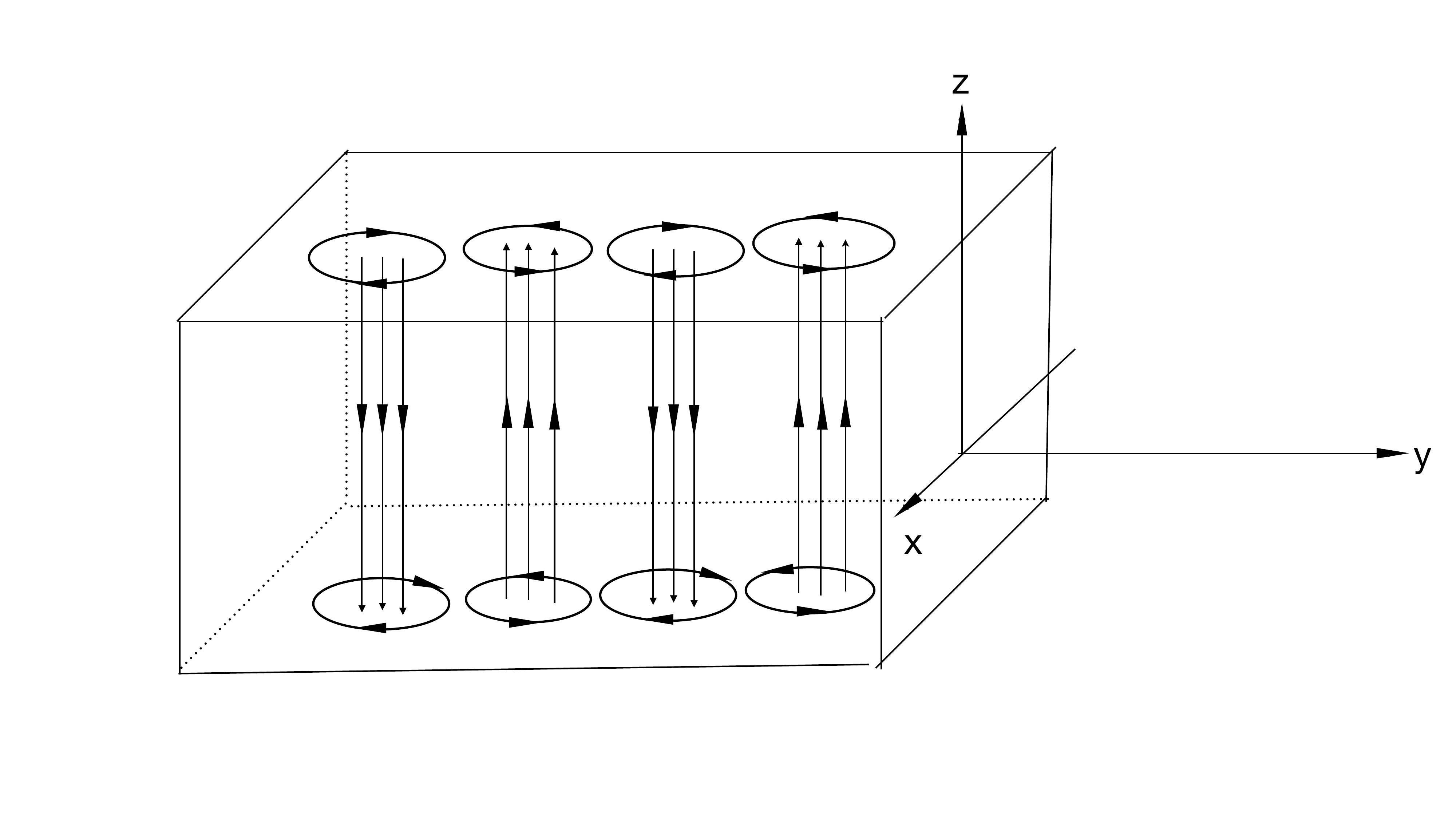} ~~
\includegraphics[angle=0,width=8cm]{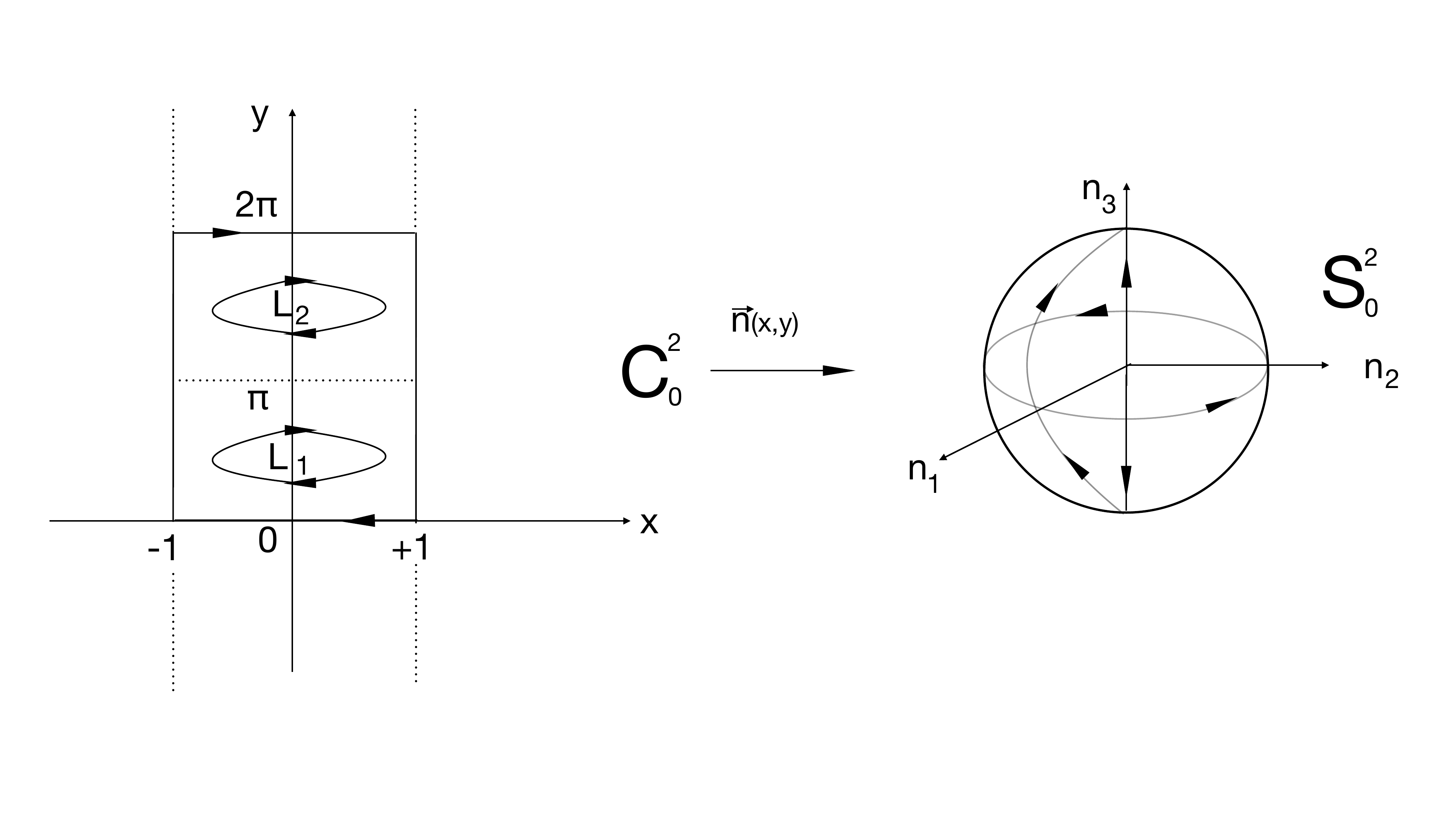} 
\centering
\caption{The figure demonstrates a finite part of an infinite wall of finite thickness ${2\over a}$ in the direction of the $x$ axis of the solution (\ref{polsol}), (\ref{magsheet}). It is  filled by  parallel chromomagnetic fluxes of opposite orientation (see Appendix B for details). Each chromomagnetic flux tube cell of the square area ${2\over a}  {  \pi \over b}$ carries the  flux ${4 \pi \over g}$.  The circuits in the left figure show the flow of the conserved current   $J^a_{k}=g \epsilon^{abc} A^b_{j} G^c_{ik} $ and the vertical arrows show the vorticity directions $\omega^a_i = \epsilon_{ijk} \partial_j J^a_k $ (\ref{vorticity}).   In the right figure the unit vector $n^a =(\sqrt{1-x^2} \cos y, \sqrt{1-x^2} \sin y,x )$ defines the map of a unit cell ${\bf C}^2: x\in (-1,1); y \in (0,2\pi)$ to a sphere  ${\bf S}^2$.   }
\label{fig1} 
\end{figure}
Let us consider solutions through which one can expose the essential properties of the general solution.  To obtain a particular solution in an explicit form we have to choose the function $\theta(X)$. Considering  $\theta(X)=\arcsin(\sqrt{1- (a\cdot x)^2})$ we obtain a "polynomial"  solution  \cite{Savvidy:2024sv, Savvidy:2024ppd,Savvidy:2024xbe}:
\be\label{polsol}
n^a(x)= \{ \sqrt{1-(a\cdot x)^2} \cos(b\cdot x),~ \sqrt{1-(a\cdot x)^2} \sin(b\cdot x),~ (a\cdot x)  \},
\ee 
which represents a  magnetic flux tubes of a finite thickness $  2/\vert a \vert $, and the corresponding gauge field (\ref{choansatzint}) has the following form:
\beqa\label{magsheet} 
A^{a}_{i}(x,y) = {1\over g} \left\{
\begin{array}{llll}   
 {1 \over \sqrt{1-(a x)^2} }\Big(  a \sin b y - gH y  (1-(a x)^2) \cos b y ,\\
 -   a \cos b y  - gH y  (1-(a x)^2)\sin b y, -g H a   x  y \sqrt{1-(a x)^2}\Big)& \\
b \sqrt{1-(a x)^2} \Big(-a   x \cos b y , - a    x \sin b y ,  \sqrt{1-(a x)^2}\Big)&\\
(0,0,0),~~~~~~~~~~~~~~~~~~~~~~~~~~~~~~~~~~~~~~~~~~~~~~~~~~~~~~~~~~~~~~~~~~~~~~~~ (a  x)^2 <  1, &
\end{array} \right.   
\eeqa 
where $\vec{a}=(a,0,0)$, $\vec{b}=(0,b,0)$,  $B_1=-H y$ and $A^{a}_{\mu} =0$ when $ (a  x)^2 > 1 \nn$.  The non-zero component of the field-strength tensor is
\be\label{inducedmag1}
G^a_{12}(x,y)= {gH-a b \over g} \Big( \sqrt{1-(a x)^2}  \cos  b y  ,~ \sqrt{1-(a x)^2}   \sin  b y ,~ a  x   \Big).
\ee
The distribution of currents that support the solution geometry are obtained by calculating the conserved current:
\be
J^a_{\mu} = g \epsilon^{abc} A^b_{\nu} G^c_{\nu\mu},~~~~~~~~~\partial_{\mu} J^a_{\mu}=0.
\ee
The non-vanishing components of the chromoelectric current supporting the chromomagnetic field are:
\beqa\label{electiccur}
&J^1_{1}= { b(g H- a b)\over g} \sqrt{1-(a x)^2} \sin b y, ~~~~~~~&J^1_{2}= -{a^2 (g H- a b) x \over g} {\cos b y \over \sqrt{1-(a x)^2} },\nn\\
&J^2_{1}= -{b(g H- a b)\over g} \sqrt{1-(a x)^2} \cos b y, ~~~~~~~&J^2_{2}= -{a^2(g H- a b) x \over g} {\sin b y \over \sqrt{1-(a x)^2} },\nn\\
&J^3_{1}= 0, ~~~~~~~~&J^3_{2}= {a(g H- a b)  \over g}.  
\eeqa  
One can check that $ \partial_{\mu} J^a_{\mu}=    \partial_1 J^a_1 + \partial_2 J^a_2 =0 $. The non-zero component of the current vorticity $\omega^a_i = \epsilon_{ijk} \partial_j J^a_k $  is
\be\label{vorticity}
\omega^a_3 ={1\over g}  { (a b - gH)(a^2 +b^2(1- a^2 x^2)^2 ) \over (1-(a x)^2)^{3/2}  }  \Big( \cos b y,  \sin b y,0 \Big),~~~ (a  x)^2 <  1.
\ee
\begin{figure}
 \centering
\includegraphics[angle=0.2,width=11cm]{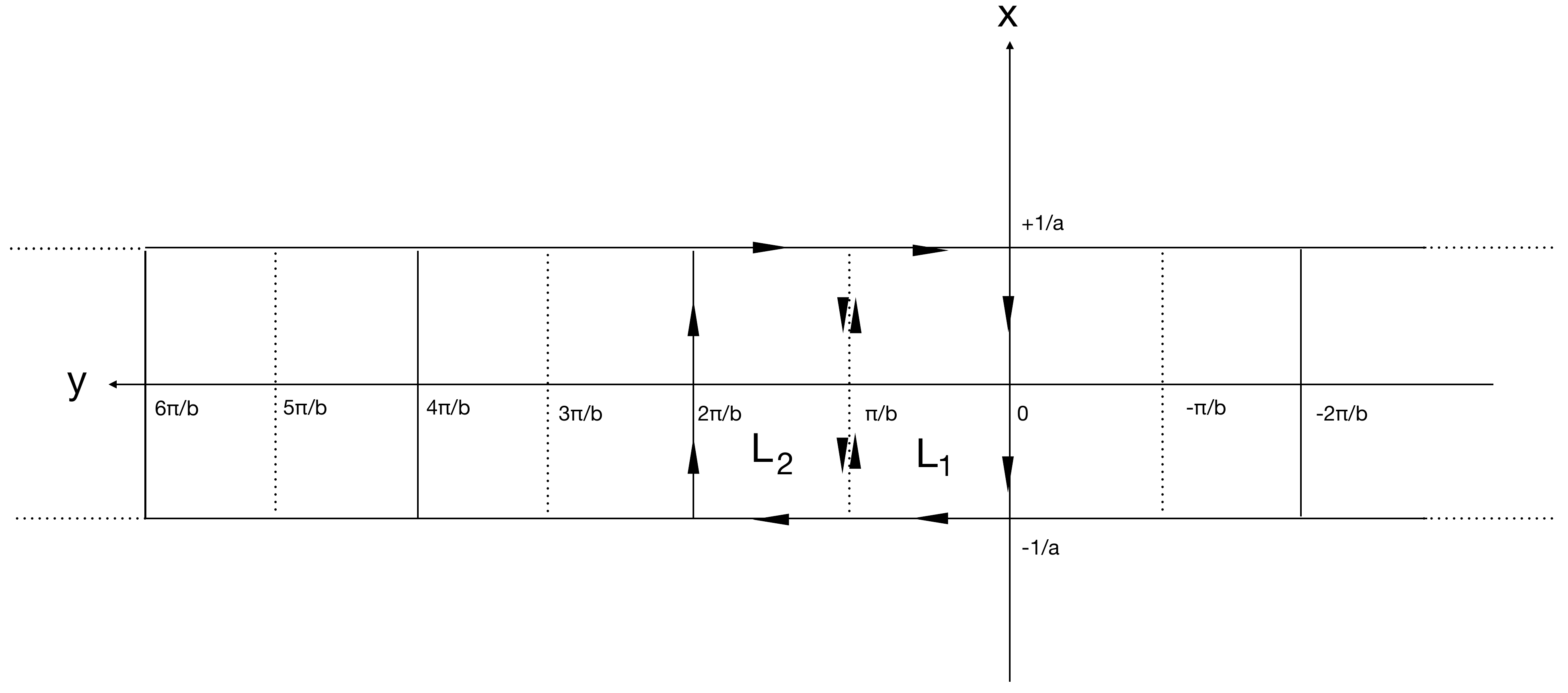} ~~
\centering
\caption{The figure demonstrates  the geometry of a chromomagnetic flux tube of finite thickness ${2\over a}$ in the direction of the $x$ axis and infinite in $y$ and $z$ axis. It is a section of the solution (\ref{polsol}), (\ref{magsheet}), (\ref{magsheet1} ) by the plane $(x,y,0)$ when $g H =0$. A space is  filled by  parallel chromomagnetic fluxes of opposite orientation. Each chromomagnetic flux tube cell of the square area ${2\over a}  {  \pi \over b}$ carries the  flux ${2 \pi \over g}$. $L_1, L_2$ are the integration contours in the operator $A(L)$ (\ref{magflux}). }
\label{fig11} 
\end{figure}
It is  singular at the location of the wall boundaries  $x=\pm 1/a$.  There is no energy flow from the magnetic flux wall in the direction transversal to the wall boundaries  because the Poynting vector vanishes, $\vec{E^a} \times \vec{H^a} =0$.  This solution is similar to the superposition of the Nielsen-Olesen magnetic flux tubes and is supported without presence of any Higgs field (see Fig.\ref{fig1},\ref{fig11}).  The magnetic flux is defined by the nonlocal gauge-invariant operator\footnote{$2 A(L)$ is a character of the SU(2) representations $\chi_j = {\sin(j+1/2) \Phi  \over \sin(\Phi/2) }$ and for $j=1/2$ it is $ \chi_{1/2} = 2\cos(\Phi/2)$. } \cite{tHooft:1981bkw, tHooft:1979rtg,tHooft:1980xss}:
\be\label{magflux}
A(L)={1\over 2} Tr P \exp{(i  g  \oint_L \hat{A}_k d x^k)} =  \cos{\Big({1\over 2}  g  \Phi\Big)}.
\ee
Let us consider a closed loop $L$ surrounding an oriented magnetic flux tube cell of the square area $  { 2\pi \over a b}$ on the $(x,y)$ plane of the solution  (\ref{magsheet}) when $gH=0$ \footnote{ The gauge-invariant  flux defined in (\ref{magflux}) is not a strictly additive quantity. } (see Fig.\ref{fig11} and Appendix B).  For the closed  contour $L_1: y=0, x\in (1/a,-1/a)$;~ $ y=\pi/b,  x\in (-1/a,1/a)$  of the square area $ { 2 \pi \over a b}$ in   Fig.\ref{fig1} the  phase factor is
\be\label{magflux123}
 \oint_{L_1 } \hat{A}_{\mu} d x_{\mu}=  \frac{ \pi}{g} \sigma_2, 
\ee
and the magnetic flux through the contour  $L_1$  is  $\Phi(L_1)={2\pi \over g}$ and  $A(L_1)=-1$.   Considering the alternative contour $L_2: y=\pi/b, x\in (1/a,-1/a)$,~ $ y=2\pi/b,  x\in (-1/a,1/a)$  of the area $ { 2 \pi \over a b}$ we will obtain  the negative phase factor:
\be
 \oint_{ L_2} \hat{A}_{\mu} d x_{\mu}=  - \frac{ \pi}{g}  \sigma_2, ~~~~~~~~~~A(L_2)=-1.
\ee
 The chromomagnetic fluxes have  {\it opposite orientations} in these cells. This fact can be illustrated by computing the total flux through the  union of these two  cells $L_1\cup L_2$: $x \in (-1/a,1/a), y \in (0, 2\pi/b)$ of the area $\frac{4 \pi}{a b}$,    which vanishes:
\be
 \oint_{L_1 \cup L_2} \hat{A}_{\mu} d x_{\mu}= 0,~~~~~~~~\Phi(L_1\cup L_2)=0,~~~~~A(L_1\cup L_2)=1.
\ee
This structure of the alternating fluxes periodically repeats itself in the direction of the $y$ axis\footnote{The magnetic flux induced by the constant Abelian field $A_1 = - H y$  through the area  $ { 2 \pi \over a b}$ is
\be\label{fluxdef1234c}
A(L )={1\over 2} Tr P \exp{(i  g \oint_{L} \hat{A}_k d x_k)} = {1\over 2} Tr e^{- i g H \frac{ 2\pi}{a b}  \frac{\sigma_1}{2}}  =\cos{\Big({ \pi \over a b }  g H \Big)}.
\ee
 }.

\begin{figure}
 \centering
\includegraphics[angle=0,width=3cm]{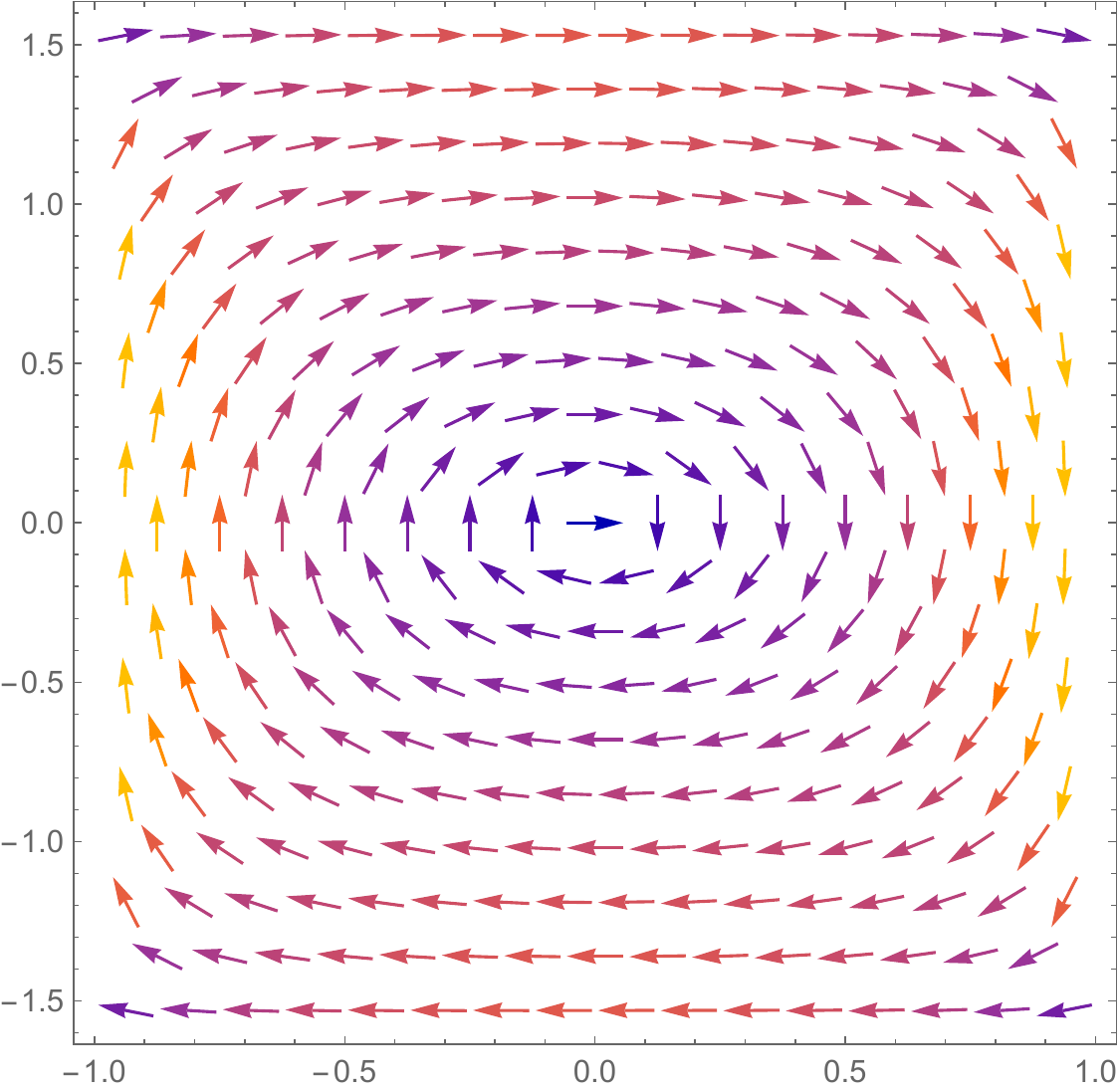} ~~
\includegraphics[angle=0,width=3cm]{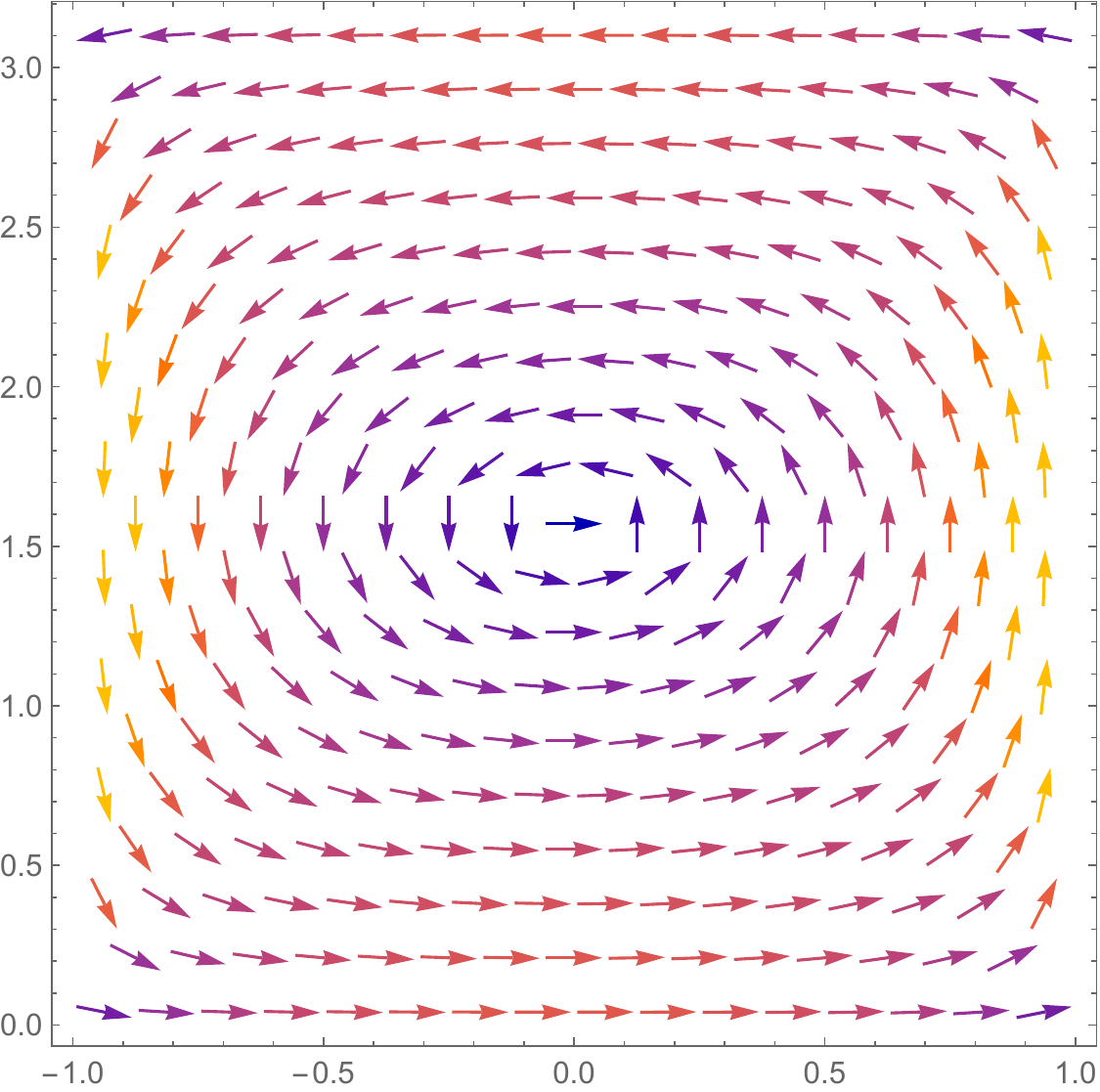} 
\centering
\caption{The flow of the chromomagnetic currents $(J^a_1(x,y),J^a_2(x,y))$ (\ref{eleccurr}) in the plane normal to the $z$ axis in two neighbouring cells $L1$ and $L_2$ defined in Fig.(\ref{fig11}).   }
\label{fig22} 
\end{figure}

To illustrate the internal structure of the covariantly constant gauge field configurations let us turn to the Amp\'ere-Maxwell-like law in the Yang-Mills theory.  The classical Yang-Mills equation can be written in the following form:
\be
  \partial_{\nu} G^a_{\nu\mu} = g \varepsilon^{a c b} A_{\nu}^{c} G^b_{\nu\mu}, 
\ee
where the right-hand side of the equation represents a conserved "self-induced" current $J^a_{\mu}$:
\be
\partial_{\nu} G^a_{\nu\mu}= J^a_{\mu}, ~~~~~J^a_{\mu}= g \varepsilon^{a c b} A_{\nu}^{c} G^b_{\nu\mu},~~~~~~\partial_{\mu} J^a_{\mu}=0.
\ee
In the case of pure chromomagnetic field the equation will take a form similar to the Amp\'ere-Maxwell equation:
\be
\varepsilon_{ijk} \partial_{j} H^a_k = - J^a_{i}, ~~~~~J^a_{i}= g \varepsilon^{a c b} A_{j}^{c} G^b_{ji},  
\ee
where $G^a_{ij} = \varepsilon_{ijk} H^a_k$.  In the vector notation the equation will take the following form:
\be
\vec{\nabla} \times  \vec{H}_a = - \vec{J}_a. 
\ee
In its integral form the equation defines the circulation of the chromomagnetic field $\vec{H}_a $ around the contour $L$ in terms of the total flux of the chromoelectric current $\vec{J}_a$ through the surface $\Sigma$:
\be\label{inteqAmpMax}
  \oint_{L}  \vec{H}_a   d \vec{x} = - \oint_{\Sigma}  \vec{J}_a d \vec{\sigma},
\ee
where $L = \partial \Sigma $ is the boundary of the two-dimensional surface $\Sigma$. We can now illustrate how the chromomagnetic field and the chromoelectric current interact creating the  flux-tube solution (\ref{magsheet}). The nonzero component of the chromomagntic field  (\ref{inducedmag1}) is  ($H=0$)
\be
H^a_{3}= -{a b \over g} \Big( \sqrt{1-(a x)^2}  \cos  b y  ,~ \sqrt{1-(a x)^2}   \sin  b y ,~ a  x   \Big),
\ee
and it is created by the chromoelectric current (\ref{electiccur}), which has the following non-vanishing components:
\beqa\label{eleccurr}
 \frac{\partial H^a_3}{\partial y}= - J^a_{1}, ~~~~~~~~~~~~~~~~~~~ \frac{\partial H^a_3}{\partial x}= J^a_{2},~~~~~~~~~~~~~~~~~~~~~~~~~~~~~~~\\
 J^a_{1}={  a b^2 \over g} \sqrt{1-(a x)^2}  \Big(- \sin b y, \cos b y, 0\Big),   ~~~~~
J^a_{2}= {  a^2 b  \over g}   \Big(  {a x \cos b y \over \sqrt{1-(a x)^2} },    { a x \sin b y \over \sqrt{1-(a x)^2} }, - 1\Big).\nn
\eeqa  
The integral equation (\ref{inteqAmpMax}) takes the following form:
\beqa\label{intwall}
 & \oint_{L}   ( \sqrt{1-(a x)^2}  \cos b y, \sqrt{1-(a x)^2} \sin by, a x  )    d z =\\
&   b  \oint_{\Sigma}  \sqrt{1-(a x)^2} ( - \sin b y,  \cos by, 0  ) dy dz + 
a  \oint_{\Sigma} \Big( \frac{ a x \cos b y}{\sqrt{1-(a x)^2} } , \frac{ a x \sin b y}{\sqrt{1-(a x)^2} }, -1  \Big) dx dz . \nn 
\eeqa 
Let us specify the surface $\Sigma$ to be in the plane $x=0$ with the boundary $y \in [0,\pi/b]$, $z\in [0,L]$. The circulation of the chromomagnetic field will be $\oint_{L}  \vec{H}_a   d \vec{x}= (2L,0,0)$, and, as one can check by using (\ref{intwall}), it is equal to the total chromoelectric flux $- \oint_{\Sigma}  \vec{J}_a d \vec{\sigma}=(2L,0,0)$. For the surface $\Sigma$ in the plane $y=0$ and the boundary $x \in [-1/a,1/a]$, $z\in [0,L]$ one can get 
\be
\oint_{L}  \vec{H}_a   d \vec{x}= (0,0,-2L),~~~~~~\oint_{\Sigma}  \vec{J}_a d \vec{\sigma}=(0,0,-2L).
\ee
The flow of the chromoelectric current is illustrated in the Fig.(\ref{fig22}).
In the limit $a \rightarrow 0 $  the chromomagnetic field will spread all over the 3d-space and by considering $b \rightarrow \infty$  while keeping the product  $a b$ fixed will define a finite energy density solution $ \epsilon = {  a^2 b^2  \over 2 g^2}$ in the whole 3d-space. 

When $\theta(X)=\arcsin({1\over  \cosh(a \cdot x)})$, we will obtain a "hyperbolic" solution, which has the infinite width in the $x$ direction unlike the finite width of the solution (\ref{polsol}):
\be\label{hypersol}
n^a(x)= \{ {\cos((b\cdot x) \cosh^2(a\cdot x)) \over  \cosh(a\cdot x)}, { \sin((b\cdot x) \cosh^2(a\cdot x)) \over \cosh(a\cdot x)}, \tanh(a\cdot x)  \}.
\ee 
Finally,  when $\theta(X)= (a\cdot   x) $, we will obtain a  "trigonometric" solution:
\be\label{ansatz2}
n^a(\vec{x})= \{ \sin a x \cos \Big({b y \over \sin a x }\Big),~\sin a x \sin\Big({b y \over \sin a x}\Big),~ \cos a x  \}. 
\ee 
Here as well the chromomagnetic flux tubes form a periodic lattice structure distributed in space and have their fluxes oriented in the opposite directions, similar to the superposition of the Nielsen-Olesen magnetic vortices \cite{Nielsen:1973cs, Nambu:1974zg}.

{\it The importance of having exact solutions of the sourceless Yang-Mills equation lies in the fact that only in that case the vacuum polarisation and the effective Lagrangian represent the gauge-invariant physical effects}   \cite{Savvidy:1977, Batalin:1976uv, Batalin:1979jh}.  

As we already mentioned the above solution (\ref{choansatzint}), (\ref{generasolint}) can be considered as a solution of the Yang-Mills equation in the background field $F_{\mu\nu}$ that has additional non-Abelian term $S_{\mu\nu}$.  The classical solutions of the Yang-Mills equation in the constant background field (\ref{consfield}) were first considered  in \cite{Ambjorn:1978ff, Ambjorn:1980ms}, and it was found  that  in the linear approximation the excitation of the negative-mode amplitude  $W$ (see equation (\ref{negmodampli})) generates a periodic lattice structure of magnetic flux tubes.  Beyond the linear approximation  the negative-mode  amplitude $W$ was considered in \cite{Arodz:1980gh}, and it was found that there are no nontrivial solutions of the Yang-Mills equation  and $W=0$  (see Appendix D for details)\footnote{ The solution (\ref{choansatzint}), (\ref{generasolint}) of the Yang-Mills equation  is not in the subspace  of the $W$ mode.}.

The ansatz (\ref{choansatzint}), (\ref{abelean}) and its extensions were considered in \cite{ tHooft:1974kcl, Corrigan:1975zxj, Cho:1979nv,  Biran:1987ae, Faddeev:1998eq, Faddeev:2001dda, Savvidy:2024sv, Savvidy:2024ppd, Savvidy:2024xbe}.  In the first case  as the electromagnetic field-strength tensor of a magnetic monopole in the Yang-Mills-Higgs model and as a truncation of the full four-dimensional connection $A^a_{\mu}$.  The goal was to identify those field degrees of freedom in $A^a_{\mu}$ that are expected to be relevant for the description of the Abelian dominance \cite{Cho:1979nv}. 
In an alternative approach \cite{Faddeev:1998eq, Faddeev:2001dda,Kondo:2005eq}  a  reformulation of the SU(2) Yang-Mills theory was suggested in terms of the field components that are written in this orthonormal frame, and it was conjectured  that the new variables describe the theory in its infrared regime with string-like excitations  \cite{Faddeev:1996zj,Faddeev:2006bm, Shabanov:1999xy}.  In  \cite{Duan:1979ucg, Kondo:2025dtx} the authors were considering the  ansatz (\ref{abelean}) in the Yang-Mills-Higgs model with a unit vector $n^a$ associated with the adjoint scalar field $n^a = { \phi^a \over \vert \phi \vert}$.

In this article our aim is to calculate the effective  Lagrangian for the covariantly constant gauge fields  (\ref{YMeqcovint1}), (\ref{magneticsheetsolution1}).  In the next section we will review the evaluation of the effective Lagrangian in the Yang-Mills theory in the case of a constant  field (\ref{consfield}) and then will calculate the effective Lagrangian for the covariantly constant gauge fields (\ref{magneticsheetsolution1}).   In QED the polarisation of the vacuum and the effective Lagrangian were obtained in two important cases: for the constant electromagnetic field \cite{Heisenberg:1936nmg} and the plane wave solution \cite{Schwinger:1951nm}. In the Yang-Mills theory the effective Lagrangian was obtained only for the constant field (\ref{consfield}).

\section{\it Gauge invariance of the effective action for sourceless fields}

The classical action of the $SU(2)$ Yang-Mills field has the following form \cite{Yang:1954ek}:
\be\label{YMaction}
S_{Y . M .}=-\frac{1}{4} \int d^{4} x \, G_{\mu \nu}^{a} G_{\mu \nu}^{a},
\ee
where the field-strength tensor is
$
G_{\mu \nu}^{a}=\partial_{\mu} A_{\nu}^{a}-\partial_{\nu} A_{\mu}^{a}-g \varepsilon^{a b c} A_{\mu}^{b} A_{\nu}^{c}\nn
$
and $\varepsilon^{a b c}$  are structure constants of the $SU(2)$ algebra. The Yang-Mills action is invariant with respect to the infinitesimal gauge transformations:
\be\label{YMtransform}
A_{\mu}^{a} \rightarrow A_{\mu}^{a}+\nabla_{\mu}^{a b}(A) \delta \xi^{b},\nn
\ee
where
$
\nabla_{\mu}^{a b}(A)=\delta^{a b}\partial_{\mu}-g \varepsilon^{a c b} A_{\mu}^{c} 
$
is the covariant derivative and has the following property:
\be\label{nablacommut}
\left[\nabla_{\mu}, \nabla_{\nu}\right] = -g \hat{G}_{\mu \nu},
\ee
where
 $\hat{G}_{\mu \nu}^{a b} = \varepsilon^{a c b} G_{\mu \nu}^{c}$. 
 The one-loop contribution to the effective action $\overline{\Gamma}[A, \bar{A}]$  in the background gauge is \cite{Savvidy:1977, Batalin:1976uv,  Savvidy:2019grj}
\beqa\label{oneloopYMaction}
\overline{\Gamma}^{(1)}[A, \bar{A}] & = S_{\alpha}[A, \bar{A}] + \frac{i}{2} \operatorname{Sp} 
\ln \left[\frac{\delta^{2} S_{\alpha}[A, \bar{A}]} {\delta A \delta A}\right]  - i \operatorname{Sp} \ln \left[\nabla_{\mu}(\bar{A}) \nabla_{\mu}(A)\right], 
\eeqa
where \(\operatorname{Sp} = \operatorname{tr} \operatorname{\hat tr} \int d^{4} x\) is the trace over the Lorentz and internal indices and the integration is over the four-dimensional space-time.  The gauge-fixed action $S_{\alpha}$ has the following form:
\be\label{gaugeYMaction}
S_{\alpha}[A, \bar{A}] = S_{YM}[A] - \frac{\alpha}{2} \int d^{4} x \left[\nabla_{\mu}^{a b}(\bar A)(A - \bar{A})_{\mu}^{b}\right]^{2},
\ee
where $\alpha$ is a gauge parameter and we are  considering  the extension of the background gauge   \cite{tHooft:1976snw, Abbott:1980hw, Honerkamp:1972fd, Kallosh:1974yh, Arefeva:1974jv,Sarkar:1974ni,Kluberg-Stern:1974nmx, Kluberg-Stern:1975ebk,tHooft:1973bhk}. The  field $\bar{A} $ is considered as "external"  in all functional derivatives,  and it should be taken equal to $A$ thereafter \cite{Savvidy:1977, Batalin:1976uv}:
\be
\Gamma[A] = \overline{\Gamma}[A, \bar{A}] \vert_{\bar{A}=A}. 
\ee 
Our aim is to investgate  the effective action $ \Gamma[A]$ for the gauge fields $A$ that are the solutions of the sourceless Yang-Mills equation 
\be\label{YMequation}
\nabla^{ab}_{\mu} G_{\mu \nu}^{b} = 0
\ee
and in particular for the covarianty constant gauge fields (\ref{YMeqcovint1}). Below we will  prove the gauge invariance of the effective action for sourceless gauge fields   \cite{Savvidy:1977, Batalin:1976uv, Batalin:1979jh}. The gauge invariance of the  effective action $ \Gamma[A]$  and its independence from the gauge parameter $\alpha$ will be proved by using the  Slavnov-Taylor-like identity \cite{Slavnov:1972fg,Taylor:1971ff}.  

By calculating the second functional derivative of the action $S_{\alpha}[A, \bar{A}]$ (\ref{gaugeYMaction}) and taking $\bar{A} = A$, one can get
\be\label{oneloopaction}
\Gamma[A] = S_{YM} +W_{YM}^{(1)} + W_{FP}^{(1)},
\ee
where
\be\label{oneloopapprox}
W_{YM}^{(1)} = \frac{i}{2} \operatorname{Sp} \ln \left[ H(\alpha) \right], ~~~~~~~W_{FP}^{(1)} = -i \operatorname{Sp} \ln \left[ H_{0} \right],
\ee
and 
\beqa\label{vectorghosthamilt}
H_{\mu\nu}(\alpha) = g_{\mu\nu} \nabla_{\sigma} \nabla_{\sigma} - 2 g \hat{G}_{\mu\nu} + (\alpha-1) \nabla_{\mu} \nabla_{\nu}, ~~~~~~~~~
H_{0} = \nabla_{\sigma} \nabla_{\sigma}. \label{ghosth}
\eeqa
The Green functions for the Yang-Mills and ghost fields  in the background field are defined by the following operator equations:
\be\label{greenfunctions}
H(\alpha) \Delta = -1,  ~~~
H_{0} \mathcal{D} = -1. 
\ee
By using the Heisenberg-Euler-Fock-Schwinger proper time parametrisation \cite{Heisenberg:1936nmg, Fock:1937ab, Fock:1937dy, Nambu:1950rs, Schwinger:1951nm,Avramidi:1990je} one can represent the one-loop effective action in the following form \cite{Savvidy:1977, Batalin:1976uv, Savvidy:2019grj}:
\beqa\label{propertimerep}
W_{YM}^{(1)}(\alpha) = -\frac{i}{2} \int \frac{ds}{s} \operatorname{Sp} \left[ e^{-i H(\alpha) s} \right],~~~~
W_{FP}^{(1)} = i \int \frac{ds}{s} \operatorname{Sp} \left[ e^{-i H_{0} s} \right],
\eeqa
and the Green functions  as
\beqa\label{greenfunctions1}
\Delta(\alpha) = -i \int ds \exp \left\{ -i H(\alpha)  s \right\},~~~~~~~~
\mathcal{D} = -i \int ds \exp \left\{ -i H_0 s \right\}.
\eeqa
We have to define the dependence of the effective action  $W^{(1)}_{YM}(\alpha)$  and of the Green function  $\Delta(\alpha)$  on the gauge  parameter $\alpha$ for the class of the sourceless gauge fields (\ref{YMequation}).  This dependence  can be investigated by using the fundamental relation for the operator $H_{\mu\nu}(\alpha)$.   Acting on   $H_{\mu\nu}(\alpha)$ (\ref{vectorghosthamilt}) by the operator $\nabla_\mu$  from the left-hand side   one can get
\begin{equation}
\nabla_\mu H_{\mu\nu}(\alpha) = \alpha H_0 \nabla_\nu - g \left[ \nabla_\mu, \hat{G}_{\mu\nu} \right],
\end{equation}
and because 
\be\label{eqmoution}
\left[ \nabla_\mu,  \hat{G}_{\mu\nu} \right] \equiv \widehat{ \nabla_\mu  G}_{\mu\nu} =0,
\ee
for the sourceless gauge fields (\ref{YMequation})  
\begin{equation}\label{fundequation}
\nabla_\mu H_{\mu\nu}(\alpha) = \alpha H_0 \nabla_\nu.
\end{equation}
This relation is a direct consequence of the gauge invariance of the Yang-Mills action  $S_{YM}$. Indeed, from   (\ref{YMaction}) we have
$
S_{YM}[A + \delta A] = S_{YM}[A],  
$
so that
\begin{equation}
\nabla_\mu^{ab} \frac{\delta S_{YM}}{\delta A_\mu^b(x)} = 0. \nonumber
\end{equation}
Calculating  the functional derivative over $A_\nu^d$ we obtain that
\begin{equation}
-g \varepsilon^{adb} \delta(x-y) \frac{\delta S_{YM}}{\delta A_\nu^b(x)} + \nabla_\mu^{ab}(x) \frac{\delta^2 S_{YM}}{\delta A_\mu^b(x) \delta A_\nu^d (y)} = 0.  \nonumber
\end{equation}
For the sourceless fields (\ref{YMequation})  $  \frac{\delta S_{YM}}{\delta A_\nu^b(x)}=\nabla_\mu^{bc} G^c_{\mu\nu} = -J^b_{\nu} = 0$ we will have 
$$ 
\nabla_\mu^{ab}(x) \frac{\delta^2 S_{YM}}{\delta A_\mu^b(x) \delta A_\nu^d (y)} =\nabla_\mu^{ab} H^{bd}_{\mu\nu}(0) =0,
$$ which  leads to the relation (\ref{fundequation}). Now, using the relation (\ref{fundequation}) one can find the dependence on $\alpha$ of the operator  $U_\alpha = \exp \{-i H(\alpha) s\}$ and therefore of the one-loop effective action $W_{Y.M}^{(1)}(\alpha)$ (\ref{propertimerep}) and of the Green function  $\Delta(\alpha)$ (\ref{greenfunctions1}).  By using the variational property of the determinant
$
\delta \ln \det X = \delta \operatorname{Sp} \ln X = \operatorname{Sp} X^{-1} \delta X  
$
one can get from (\ref{oneloopapprox}), (\ref{vectorghosthamilt}) and (\ref{greenfunctions}) that 
\be\label{actionvariation}
\delta W^{(1)}_{YM} (\alpha) = -\frac{i}{2} \operatorname{Sp} \left\{ \Delta(\alpha) \delta H(\alpha) \right\} 
 = -\frac{i}{2} \delta \alpha \operatorname{Sp} \left\{ \Delta_{\mu\nu}(\alpha) \nabla_{\nu} \nabla_{\mu} \right\} = -\frac{i}{2} \operatorname{Sp} \left\{ \nabla_\mu \Delta_{\mu\nu}(\alpha) \nabla_\nu \right\} \delta \alpha.
\ee
To find the expression under the trace one should act by the operator  $ \nabla_\mu $ on  $H_{\mu\nu} (\alpha)$ from the right and by the operator $\nabla_\nu $ from the left in the  formula (\ref{greenfunctions})
\be\label{greenfunctions2}
H_{\mu\nu} (\alpha)  \Delta_{\nu\lambda}(\alpha) = -g_{\mu\lambda} 
\ee
and then using the equation (\ref{fundequation}),
\be\label{inverse}
\alpha H_0 \nabla_\mu \Delta_{\mu\nu}(\alpha) = - \nabla_\nu,
\ee
we get that
\be
\nabla_{\mu} \Delta_{\mu\nu}(\alpha) \nabla_{\nu} = -\frac{1}{\alpha}.
\ee
Integrating the equation (\ref{actionvariation}) we obtain the explicit dependence of the one-loop effective action on the gauge parameter $\alpha$:
\be\label{alphadepen}
W_{YM}^{(1)}(\alpha) = W_{YM}^{(1)}(1) + \frac{i}{2} \ln \alpha  \operatorname{Sp}  \mathbbm{1}.
\ee
It follows that up to the trivial term $ \frac{i}{2} \ln \alpha  \operatorname{Sp}  \mathbbm{1} $, which does not depend on the gauge field, the one-loop effective action $\Gamma$ is a  gauge-invariant functional and is an $\alpha$-independent functional for the sourceless gauge fields  (\ref{YMequation}). Therefore  we have \cite{Savvidy:1977, Batalin:1976uv, Savvidy:2019grj}
\be\label{finoneloopeffact}
W_{YM}^{(1)} = -\frac{i}{2} \int_{0}^{\infty} \frac{ds}{s} \, \operatorname{Sp}  \, U(s), ~~~~~~~~
W_{FP}^{(1)} = i \int_{0}^{\infty} \frac{ds}{s} \,\operatorname{Sp}  \, U_0(s),
\ee
where 
\be\label{foperats}
U(s) = e^{-iH(1)s},~~~~~~~~~
U_0(s) = e^{-iH_0s},
\ee
and 
\beqa\label{vectorhamilt1}
H_{\mu\nu}(1) &=& g_{\mu\nu} \nabla_{\sigma} \nabla_{\sigma} - 2 g \hat{G}_{\mu\nu}, ~~~~~~~~H_{0} = \nabla_{\sigma} \nabla_{\sigma}.
\eeqa
In a similar way we can find the $\alpha$-dependence of the  propagator $\Delta_{\mu\nu}(\alpha)$. Representing the (\ref{greenfunctions}), (\ref{greenfunctions2})  in the following form:
\be
H_{\mu\nu}(1) \Delta_{\nu\lambda}(\alpha) + (\alpha-1) \nabla_{\mu} \nabla_{\nu} \Delta_{\nu\lambda}(\alpha) = -g_{\mu\lambda}
\ee
and using  the relation (\ref{inverse})
\be
H_{\mu\nu}(1) \Delta_{\nu\lambda}(\alpha) - \frac{\alpha-1}{\alpha} \nabla_{\mu} H_{0}^{-1} \nabla_\lambda = -g_{\mu\lambda}
\ee
we will find that
\be\label{gaugegreenfuntion}
\Delta(\alpha) = -\frac{1}{H(1)} \left[ 1 - \frac{\alpha-1}{\alpha} \nabla \frac{1}{ H_0} \nabla \right] 
\ee
or that  the Green function of the gauge boson in the background field has the following form:
\be
\Delta(\alpha) = \Delta(1) \left[ 1 + \frac{\alpha-1}{\alpha} \nabla \mathcal{D} \nabla \right] 
=\Delta(1) \left[ 1 + \nabla \mathcal{D} \nabla \right] - \frac{1}{\alpha} \Delta(1) \nabla \mathcal{D} \nabla =\Delta_T - \frac{1}{\alpha} \Delta_L , 
\ee
where
\be
\Delta_T= \Delta(1) \left[ 1 + \nabla \mathcal{D} \nabla \right],~~~\quad \nabla \cdot \Delta_T = 0 \,.
\ee
In the proper time representation the gauge Green function (\ref{gaugegreenfuntion}) has the following form:
\be\label{fgreenfunct}
\Delta(\alpha) = -i \int ds \, U(s) - \frac{\alpha - 1}{\alpha} \int ds dt \, U_0(s) \nabla U(t) \nabla,
\ee
and the ghost Green function (\ref{greenfunctions1}) is
\be\label{fgreenfunctghost}
\mathcal{D} = -i \int ds \, U_0(s).
\ee
The gauge invariance of the effective action for sourceless gauge fields can also be proved without reference to a loop expansion by using the Slavnov-Taylor-like identity \cite{Savvidy:1977}:
\beqa\label{batsaviden}
 \alpha\frac{\mathrm{d} \bar{\Gamma}}{\mathrm{d} \alpha} 
= \frac{1}{2} \left\langle \int d^4 x \, d^4 y\quad \frac{\delta \overline{\Gamma}}{\delta A_{\mu}^{a}(x)}  \,\nabla_\mu^{ab}(A) \mathcal{D}^{bc}(x, y) 
\nabla_\nu^{cd}(\bar{A}) (A-\bar{A})_\nu^d \right\rangle_c. 
\eeqa
It follows that on sourceless gauge fields  
\be
 \quad \frac{\delta \overline{\Gamma}}{\delta A_{\mu}^{a}(x)}= -J_\mu^a(x) =0
\ee
 and the effective action  $\bar{\Gamma}$ is a gauge invariant functional:
\be\label{actioninvar}
 \frac{\mathrm{d} \bar{\Gamma}}{\mathrm{d} \alpha} =0.
\ee 
 There is a  strong physical constraint  on any possible imaginary term  in the effective action that follows from  the expression for the vacuum persistence probability given by the formula \cite{Schwinger:1951nm}:
\be\label{probabi}
|\langle 0\vert 0 \rangle|^2=  \left| \exp \left\{ i \Gamma \right\} \right|^{2} = \exp \left\{ -2~ \mathcal{I}m ~\Gamma \right\}.  
\ee
The imaginary part of $\Gamma$ defines the decay rate of the vacuum and therefore imposes a constraint 
$$
2 ~\CI m \Gamma \geq 0, 
$$ 
the probability must be less or equal to one.

In summary, the problem of computing the effective action in the Yang-Mills theory in the one-loop approximation and of the Green functions in a background field is defined by the formulas (\ref{oneloopaction}), (\ref{finoneloopeffact}), (\ref{foperats}), (\ref{vectorhamilt1}) and (\ref{fgreenfunct}), (\ref{fgreenfunctghost}). In this approach the computation of the effective action and of the Green functions reduces to the  calculation of the  matrix elements of the operators  $U(s)$ and $U_0(s)$ (\ref{foperats}):
\be\label{matrixelement}
(x' |U(s)| x'') =(x'(s)|x''(0)).
\ee    
The matrix elements of the operators  $U(s)$ and $U_0(s)$ can be computed by three alternative methods \cite{Savvidy:1977, Batalin:1976uv}. In the first method suggested by Schwinger in QED one can consider the  operators $ H_{\mu\nu}(1)$  and $ H_0$  as Hamiltonians of a  "particle" moving in a background field with "particle"  space-time  coordinates $x_{\mu}(s)$  depending on the proper time $s$ \cite{Schwinger:1951nm}. The corresponding equation of motion in the operator form can be written by using the Heisenberg representation.  By introducing the "momentum" operator $ \Pi_{\mu} = i \nabla_{\mu} $ and using the commutation relation (\ref{nablacommut}) we can obtain the equation of motion for the Hamiltonian $H(0)$ \cite{Savvidy:1977}:
\beqa\label{functequation}
\frac{d x_{\mu}}{d s} &=& -i \left[ x_{\mu}, H_0 \right] = 2 \Pi_{\mu} \nn\\
\frac{d \Pi_{\mu}}{d s} &=& -i \left[ \Pi_{\mu}, H_0 \right] = i g \left( \hat{G}_{\mu \nu} \Pi_{\nu}  +   \Pi_{\nu} \hat{G}_{\mu \nu} \right)\nn\\
&=& 2 i g \hat{G}_{\mu \nu} \Pi_{\nu} + i g \left[ \Pi_{\nu} , \hat{G}_{\mu \nu} \right].
\eeqa
The matrix elements are defined by the linear equations:
\beqa\label{matrixelementsequation}
i \partial_s (x'(s) |x''(0)) &=& (x'(s) | H_0 | x''(0)), \nn\\
(i \partial'_{\mu} - i g \hat{A}_{\mu}(x')) (x'(s) | x''(0)) &=& (x'(s) | \Pi_{\mu}(s) | x''(0)),\\
 (-i \partial''_{\mu} - i g \hat{A}_{\mu}(x''))(x'(s) | x''(0)) &=& (x'(s) | \Pi_{\mu}(0) | x''(0)),\nn
\eeqa
together with the boundary condition 
\be
(x'(s) | x''(0))_{ab} \underset{s \to 0}{=} \delta(x' - x'') \delta_{ab}.
\ee
For the sourceless fields (\ref{eqmoution}) the second term in (\ref{functequation}) is equal to zero. The equation of motion for the Hamiltonian  $H(1)$ can be obtained in a similar way. In the second method the matrix elements are computed by using the path-integral representation \cite{Batalin:1970it, Savvidy:1977}:
\begin{equation}\label{functequation1}
(x' |U_{0}(s)| x'') =(x'(s)|x''(0)) = \mathcal{N}^{-1} \int \mathcal{D}t_{\mu}(s) \exp \left\{ -i \int_{0}^{s} t_{\mu}(s') t_{\mu}(s') ds' + \right. \nonumber
\end{equation}
\begin{equation}
+\left. 2g \int_{0}^{s} ds' t_{\mu}(s') \cdot \hat{A}_{\mu}(x' - 2 \int_{s'}^{s} t(\xi) d\xi) \cdot \delta(x' - x'' - 2 \int_{0}^{s} t(\xi) d\xi) \right\},
\end{equation}
and in the third method one should find  the eigenvalues of the Hamiltonian  operators $H(1)$ and $H_0$ and calculate the determinant as a product of the eigenvalues as it was originally developed by Heisenberg and Euler \cite{Heisenberg:1936nmg} and also was used by 't Hooft in his computation of the vacuum polarisation effects induced by the instanton solution \cite{tHooft:1976snw}.

The results obtained for   $W^{(1)}$, $\Delta_{\mu\nu}$ and $ \mathcal{D}$ are valid for arbitrary sourceless gauge fields  (\ref{YMequation}) and for the covariantly constant gauge fields  (\ref{YMeqcovint1}) as well.  {\it The importance of having the exact solutions of the sourceless Yang-Mills equation lies in the fact that only in that case quantum effects and vacuum polarisation can be considered as gauge-invariant physical effects   (\ref{alphadepen}) and (\ref{finoneloopeffact}) } \cite{Savvidy:1977, Batalin:1976uv, Batalin:1979jh}.  
 
We are interested to investigate the vacuum polarisation and the effective Lagrangian for the covariantly constant gauge fields (\ref{choansatzint}), (\ref{generasolint})  and investigate their physical properties. In the next section we will review the computation of the effective action for constant field (\ref{consfield1}) and then for the  gauge fields (\ref{choansatzint}), (\ref{generasolint}).

\section{\it Effective action for constant gauge field}

Now that the properties of the covariantly constant gauge fields are quite well understood, the next step is to calculate the matrix element $(x'(s) | x''(0))$ in (\ref{foperats}) and  (\ref{matrixelementsequation})  for these fields. The equation of covariantly constant fields (\ref{YMeqcovint1}) rewritten in the alternative form,
\be
\left[ \nabla_{\rho} \hat{G}_{\mu\nu} \right] = 0,
\ee
leads to the important factorisation of the operator (\ref{foperats}):
\be\label{factorization}
U(s) = \exp \left\{ 2 i g \hat{G} s \right\} U_{0}(s).
\ee
The relation (\ref{factorization}) reduces the computation of the matrix elements of (\ref{foperats})  to the computation of the matrix elements of the operator $ U_{0}(s)$. For that one should solve the system of operator equations (\ref{functequation})  or  calculate the path integral (\ref{functequation1}) in a constant field or find the spectrum of these  operators.  The details can be found in \cite{Savvidy:1977, Batalin:1976uv, Savvidy:2019grj}. Here we will present only the final expression  
\begin{equation}\label{mainmatrixelement}
(x'(s) | x''(0) ) = -\frac{i} {(4\pi s)^2}  \exp{ \left\{-\frac{i}{4}(x'-x'') \hat{K}(s) (x'-x'')  
+ \frac{i}{2} x' \hat{N} x'' - \hat{L}(s) \right\} }, 
\end{equation}
where 
\begin{equation}
 \hat{N} = ig\hat{C}, \nonumber
 \end{equation}
\begin{equation}
\hat{K}(s) = \hat{N} \operatorname{cth}(\hat{N}s), \nonumber
\end{equation}
\begin{equation}\label{lsoper}
\hat{L}(s) = \frac{1}{2} \operatorname{tr} \ln\left[(\hat{N}s)^{-1} \operatorname{sh}(\hat{N}s)\right]. 
\end{equation}
Having in hand the matrix element (\ref{mainmatrixelement}) one can calculate  the effective Lagrangian (\ref{oneloopaction}), (\ref{finoneloopeffact}) and the Green function (\ref{fgreenfunct}). The trace in (\ref{finoneloopeffact}) is 
$
S p=\operatorname{tr} \hat{\operatorname{tr}} \int d^{4} x,
$
and by using the matrix element (\ref{mainmatrixelement}) at the coincident points one can get 
\begin{equation}\label{roeffectivelagran}
\mathcal{L}^{(1)}  =-\frac{1}{32 \pi^{2}} \int \frac{d s}{s^{3}} \operatorname{tr} \hat{\operatorname{tr}} \exp \left\{2 \hat{N} s-\hat{L}(s)\right\} 
 +\frac{1}{16 \pi^{2}} \int \frac{d s}{s^{3}} \hat{\operatorname{tr}} \exp \left\{-\hat{L}(s)\right\},  
\end{equation}
where $\hat{\operatorname{tr}}$ is the trace over the isotopic  indices and  $\operatorname{tr}$ is over the Lorentz indices. Let us stress that the above expression (\ref{roeffectivelagran}) is valid for an arbitrary gauge group, and below we will evaluate the traces in the case of the $S U(2)$ group. In that case all isotopic matrices are functions of the matrix  $\hat{n}=n^{a} \varepsilon^{a}$. The calculation of traces in (\ref{roeffectivelagran}) can be performed by the use of the eigenvalues of the matrices $F_{\mu\nu}$ and $ \hat{n}$. The characteristic equation for the matrix $F_{\mu\nu}$ coincides with that in QED \cite{Heisenberg:1936nmg, Schwinger:1951nm}:
\be\label{lorentzeiegen}
F_{(1)}^2 = -\mathcal{F} - \left( \mathcal{F}^2 + \mathcal{G}^2 \right)^{1/2},~~~~~
F_{(2)}^2 = -\mathcal{F} + \left( \mathcal{F}^2 + \mathcal{G}^2 \right)^{1/2},
\ee
where $ \mathcal{F} = \frac{1}{4} F_{\mu\nu} F^{\mu\nu}$ and $ \mathcal{G} = \frac{1}{4} F_{\mu\nu} F^{\mu\nu}$.  The equation for the matrix $ \hat{n} = \varepsilon^{a}   n^a $ is
\be
\hat{n}^3 + \hat{n} = 0,
\ee
therefore the eigenvalues  are
\be
0, \pm i.
\ee
The trace over the Lorentz indices of the operator  $ \hat{L}(s)$ can be evaluated by using  (\ref{lorentzeiegen}):
\be\label{Lmatrix}
\hat{L}(s) = \ln \left[ (g F_{(1)} s i \hat{n})^{-1} \sinh (g F_{(1)} s i \hat{n}) \right]
+  \ln \left[ (g F_{(2)} s i \hat{n})^{-1} \sinh (g F_{(2)} s i \hat{n}) \right].
\ee
We have also 
\be\label{Nmatrix}
\text{tr} e^{2 \hat{N} s} = 2 \left[ \cosh (2 g F_{(1)} s i \hat{n}) + \cosh (2 g F_{(2)} s i \hat{n}) \right].
\ee
By substituting   (\ref{Lmatrix}) (\ref{Nmatrix}) into  (\ref{roeffectivelagran}) we will get
\begin{equation}
\mathcal{L}^{(1)}_{YM} = -\frac{1}{(4\pi)^2} \int \frac{ds}{s^3} ~  \hat{\operatorname{tr}}~
\frac{gF_{(1)} s i \hat{n}}{\sinh(gF_{(1)} s i \hat{n})} \cdot 
\frac{gF_{(2)} s i \hat{n}}{\sinh(gF_{(2)} s i \hat{n})} 
\end{equation}
\begin{equation}
\times \left[ \cosh(2gF_{(1)} s i  \hat{n}) + \cosh(2gF_{(2)} s i \hat{n}) - 1 \right] \nonumber,
\end{equation}
and after calculating the isotopic traces one can get 
\beqa\label{ineffectivelagra}
\mathcal{L}^{(1)}_{YM} &=& -2 \frac{1}{16\pi^2} \int \frac{ds}{s^3}  e^{-i \mu^2 s}
\frac{gF_{(1)} s  \cdot gF_{(2)} s }
{\sinh(gF_{(1)} s ) \sinh(gF_{(2)} s  )} \nonumber\\
&+& \frac{1}{4\pi^2} \int \frac{ds}{s^3} e^{-i \mu^2 s} gF_{(1)} s \cdot gF_{(2)} s \left[ \frac{\sinh gF_{(1)} s}{\sinh gF_{(2)} s} + \frac{\sinh gF_{(2)} s}{\sinh gF_{(1)} s} \right].
\eeqa
We introduced the mass parameter $\mu^2$ in order to control  the infrared singularities and  to make the integrals convergent at infinity \cite{Savvidy:1977,Batalin:1976uv}. The first integral is the  contribution of the orbital interaction term $g_{\mu\nu} \nabla_{\sigma} \nabla_{\sigma}$ in the operator $H_{\mu\nu}(1) = g_{\mu\nu} \nabla_{\sigma} \nabla_{\sigma} - 2 g \hat{G}_{\mu\nu}$.  This expression clearly demonstrates  that up to the  factor $2$ the first integral coincides with the effective Lagrangian in the scalar electrodynamics \cite{Schwinger:1951nm}.  The factor $2$ in front of the  integral is due to the increase of the phase volume through the isotopic degrees of freedom of the charged Yang-Mills gauge boson.   The second integral in (\ref{ineffectivelagra}) is associated with the contribution of the interaction of the gauge boson spin  with the background field, and it is the interaction term  $2g\hat{G}$ in the operator $H_{\mu\nu}(1)$. In the spinor electrodynamics the corresponding "particle" Hamiltonian is \cite{Schwinger:1951nm}
\be\label{qedpart}
H_{QED}= \Pi^2_{\mu} - \frac{1}{2} e \sigma_{\mu\nu} F_{\mu\nu}, ~~~~\Pi_{\mu}  =i  \nabla_{\mu}
\ee
and the term describing the interaction of the electron spin with the background field is $ \frac{1}{2} e \sigma_{\mu\nu} F_{\mu\nu}$. By using the real eigenvalues 
\be
f_1 = -i F_{(1)}, \quad f_2 = F_{(2)}, 
\ee
where $ f_1$ and  $ f_2 $ are
\be
f^2_1 = \CF + (\CF^2 +\CG^2)^{1/2}, ~~~~~~~~~f^2_{2}= -\CF + (\CF^2 +\CG^2)^{1/2},
\ee
one can observe that the second term in  the square brackets will take the form ${   \sinh (g f_{2} s)  \over \sin (g f_{1} s) } $ and the integral diverges exponentially in the infrared region at  $\vert s \vert =\infty$.  This is due to the  large contribution of the spin interaction term $2g\hat{G}$ to the effective Lagrangian that can be traced from the expression (\ref{Nmatrix}) and leads to the divergency of the proper-time integral in the infrared region.   Choosing the integration contour  in the complex  $\textcircled{s} $  plane so that the integrals will converge at large $s$, that is by the substitution $s \rightarrow -  i s$ in the first and in the third  integrals, one can represent (\ref{ineffectivelagra}) in the following form \cite{Savvidy:1977,Batalin:1976uv}:
\beqa\label{orbandspin}
&\CL^{(1)} =  {1\over 8 \pi^2} \int^{\infty}_{s_0}    {ds \over s^3} e^{- \mu^2 s}
{ (g f_{1} s)   ~(g f_{2} s)   \over \sinh (g f_{1} s)~ \sin (g f_{2} s )} +\nn\\
&+{1\over 4 \pi^2} \int^{\infty}_{ s_0}  {ds \over s} e^{-i \mu^2 s} (g f_{1} ) ~(g f_{2} )
{   \sin (g f_{1} s)  \over \sinh (g f_{2} s) } 
- {1\over 4 \pi^2} \int^{\infty}_{s_ 0}   {ds \over s} e^{- \mu^2 s} (g f_{1} ) ~(g f_{2} )  {   \sin (g f_{2} s)  \over \sinh (g f_{1} s) }.
\eeqa
 The integrals are diverging in the ultraviolet region at $s_0=0$.
In order to renormalise the Lagrangian we have to identify the ultraviolet divergences in the above integrals. These are
\beqa
&& { (g f_{1} s)   ~(g f_{2} s)   \over \sinh (g f_{1} s)~ \sin (g f_{2} s )}     = 1 -{g^2\over 6} (f^2_1 -f^2_2) s^2  +\CO(s^4) \nn\\
&&g^2 f_{1}f_{2}{  \sin (g f_{1} s )   \over \sinh (g f_{2} s)~}     = g^2 f^2_{1}   +\CO(s^2) \nn\\
&&g^2  f_{1}f_{2}{  \sin (g f_{2} s )   \over \sinh (g f_{1} s)~}     = g^2  f^2_{2}   +\CO(s^2). \nn
 \eeqa
Subtracting these terms, which are quadratic in the field-strength tensor, we will get the renormalised effective Lagrangian  \cite{Savvidy:1977, Batalin:1976uv}:
\beqa\label{YMeffLagr} 
\CL_{spin-1}^{(1)}  = &&{1\over 8 \pi^2} \int^{\infty}_{0}  {ds \over s^3} e^{- \mu^2 s}
\Big( { (g f_{1} s)   ~(g f_{2} s)   \over \sinh (g f_{1} s)~ \sin (g f_{2} s )} -1 + {1\over 6}( g s)^2 (f^2_1 -f^2_2) \Big)+\nn\\
&+&{g^2 \over 4 \pi^2} \int^{\infty}_{0} {ds \over s} e^{-i \mu^2 s}   \Big(  f_{1} f_{2} 
{   \sin (g f_{1} s)  \over \sinh (g f_{2} s) }  - f^2_1 \Big) \nn\\
&-& {g^2 \over 4 \pi^2} \int^{\infty}_{0} {ds \over s } e^{- \mu^2 s}  \Big( f_{1} f_{2} {   \sin (g f_{2} s)  \over \sinh (g f_{1} s) }  - f^2_2\Big).
\eeqa
 Now the integrals are convergent in both regions, in the infrared and in the ultraviolet one.  {\it The effective Lagrangian (\ref{YMeffLagr}) is a Lorentz- and gauge-invariant functional}.  In the forthcoming sections we will provide an alternative renormalisation scheme that cures  simultaneously the infrared and ultraviolet divergencies and is more adequate for the renormalisation of the Yang-Mills theory \cite{Savvidy:1977, Batalin:1976uv, Savvidy:2019grj}.   

In order to compare the effective Lagrangian in the Yang-Mills theory with the Heisenberg-Euler effective Lagrangian in QED let us present it in the explicit form \cite{Heisenberg:1936nmg}:
\be\label{HeiEu1}
\mathcal{L}_{spin-1/2}^{(1)} = -\frac{2}{16 \pi^2} \int_{s_0}^\infty \frac{ds}{s^3} e^{-m^2 s} \frac{e f_1 s \cdot e f_2 s}{\sinh(e f_1 s) \sin(e f_2 s)} \cdot \cosh(e f_1 s) \cos(e f_2 s),
\ee
and the effective Lagrangian in the scalar QED is
\be\label{HeiEu2}
\mathcal{L}_{spin-0}^{(1)} = \frac{1}{16 \pi^2} \int_{s_0}^\infty \frac{ds}{s^3} e^{-\mu^2 s} \frac{e f_1 s \cdot e f_2 s }{\sinh(e f_1 s) \sin(e f_2 s)}.
\ee
Let us consider  (\ref{YMeffLagr} ) in the field  $\mathcal{G} = 0$, $ \mathcal{F} > 0$ that corresponds to a pure chromomagnetic  field in an appropriate coordinate system.  In that case 
$$ f_1 = \frac{1}{2}\sqrt{G^a_{\mu\nu}G^a_{\mu\nu}}=\sqrt{\CF} \equiv H, ~~f_{2}=0, $$
and the Lagrangian is
\beqa\label{YMeffecLagrangian}
\mathcal{L}_{YM}^{(1)}(H)&=&\frac{1}{8 \pi^{2}} \int_{0}^{\infty} \frac{d s}{s^3} e^{-\mu^{2} s}\left[\frac{g H s}{\sinh g H s}-1+\frac{(g H s)^2}{6}\right]+\nonumber\\
&+&\frac{1}{4 \pi} \int_{0}^{\infty} \frac{d s}{s^3} e^{-i \mu^{2} s} g Hs \left[\sin g H s-g H s\right].
\eeqa
In the limit of the strong chromomagnetic field $g H \gg \mu^2 $ we will obtain contributions from both integrals (formula (2.3.15) on  page 35 \cite{Savvidy:1977}):
\begin{equation}\label{YMasympt}
\mathcal{L}_{YM}^{(1)}(H) \simeq \frac{(g H)^{2}}{48 \pi^{2}} \ln \frac{g H}{\mu^{2}}-\frac{(g H)^{2}}{4 \pi^{2}} \ln \frac{g H}{\mu^{2}}
=-\frac{11}{48 \pi^{2}}(g H)^{2} \ln \frac{g H}{\mu^{2}}, 
\end{equation}
where the first positive term is due to the Landau diamagnetism, the  contribution from the orbital interaction term  $g_{\mu\nu} \nabla_{\sigma} \nabla_{\sigma}$ in the operator $H_{\mu\nu}(1)$,  and the second negative term is due to the Pauli paramagnetism, the contribution of the spin-interaction term $2 g \hat{G}_{\mu\nu}$ in the $H(1)$. The contribution associated with the  vector-boson spin dominates the asymptotic behaviour of the effective Lagrangian \cite{Savvidy:1977}, which follows from (\ref{YMasympt})\footnote{The explanation of the dynamical origin of the asymptotic freedom in the Yang-Mills theory due to the spin interaction of the gauge bosons was also suggested later in \cite{Nielsen:1980sx}. }.   In QED  the corresponding asymptotics has the following form:
\begin{equation}\label{QEDasympt}
\mathcal{L}_{QED}^{(1)} \simeq \frac{(e H)^{2}}{24 \pi^{2}} \ln \frac{e H}{m^{2}} .
\end{equation}
The essential difference in the asymptotic behaviour of (\ref{YMasympt}) and (\ref{QEDasympt}) is another manifestation of the difference between the theory with the Landau pole and the asymptotically free theory.   These differences become even more transparent with the application of the renormalisation group technique to the asymptotic behaviour of the effective Lagrangians. As a result one can obtain an exact expression for the $\mathcal{L}_{YM}^{(1)}$  \cite{Savvidy:1977, Savvidy:1977as}
\be\label{savvacgg}
\mathcal{L}_{YM}^{(1)}(H)=  - \frac{H^2}{2} - \frac{11 g^2 H^2}{48 \pi^2}\Big( \ln \frac{gH}{\mu^2}- \frac{1}{2}\Big),
\ee
where $  H^2 =  \CF = \frac{1}{4} G^a_{\mu\nu}G^a_{\mu\nu} > 0$ and $\mathcal{G} = 0 $. 

For the Green functions  $ \Delta$ and  $ \mathcal{D}$ we will have 
\beqa
\Delta(x',x'') &=&S \cdot \Delta(x'-x''), ~~~~~~~~~~~~~~~\quad \mathcal{D}(x',x'') = S \cdot \mathcal{D}(x'-x'') \nn \\
\Delta(x'-x'') &=& -\frac{1}{(4\pi)^{2}} \int \frac{\mathrm{d}s}{s^{2}} U_{s}(x'-x''), ~~~~~
\mathcal{D}(x'-x'') = -\frac{1}{(4\pi)^{2}} \int \frac{\mathrm{d}s}{s^{2}} U_{0s}(x'-x''), \nn
\eeqa
where
\begin{equation}
S= \exp \left\{ g \int^{x''}_{x'} \hat{A}_{\mu}(x) \mathrm{d}x_{\mu} \right\}  = \exp \left\{ \frac{i}{2}   x'  \hat{N} x''  \right\}
\end{equation}
is the non-diagonal phase factor of the Green functions and 
\begin{equation}
U_{s}(z) =  \exp \left\{ -\frac{i}{4} z \hat{K} z - \hat{L}(s) + 2 \hat{N} s \right\}. 
\end{equation}
 Any function of the matrix $\hat{n}$ can be represented in the form
$
\mathcal{M}^{a b}(\hat{n}) = A \delta^{a b} + B \hat{n}^{a} \hat{n}^{b} + C \hat{n}^{a b},
$
therefore for the operators $\hat{K}$ and  $\hat{L}$ we will obtain:
\begin{equation}
\hat{K}(s) = \begin{pmatrix}
K(s) & 0 & 0 \\
0& K(s) & 0 \\
0 & 0 & \frac{1}{s}
\end{pmatrix},~~~~~~
K(s) = g F \coth g F s, 
\end{equation}
\begin{equation}
\hat{L}(s) = \begin{pmatrix}
L(s) & 0 & 0 \\
0 & L(s) & 0 \\
0 & 0 & 0
\end{pmatrix}, ~~~~~
L(s) = \frac{1}{2} \operatorname{tr} \ln \left[ (g F s)^{-1} \sinh (g F s) \right], 
\end{equation}
and for the  $U_{s}^{a b}$ and $U_{0s}^{a b}$ we have:
\beqa
U_{s}^{a b}(z) = \exp \left\{ -\frac{i}{4} z K(s) z  - L(s) \right\}  \Big( ( \delta^{a b} -n^{a} n^{b} ) \cosh 2 g F s + ( i \hat{n})^{a b} \sinh 2 g F s \Big) +\nn\\
+ n^{a} n^{b} \exp \left\{ -\frac{i z^2}{4 s} \right\}, \nonumber
\eeqa
\begin{equation}
U_{0s}^{a b}(z) = \exp\left\{-\frac{i}{4}zK(s)z - L(s)\right\} ( \delta^{a b} -n^{a} n^{b} )+
 n^{a} n^{b} \exp\left\{-\frac{i}{4}\frac{z^2}{s}\right\}.
\end{equation}
These expressions for the Green functions in a background field are important ingredients in the computation of the two- and higher-loop effective Lagrangian \cite{Savvidy:1977, Batalin:1978gt}.

\section{\it Imaginary parts of the effective action  } 

The significance of the presence/absence of the imaginary parts in the effective Lagrangian is connected with the fact that they define the quantum-mechanical stability of the sourceless field configurations. In our regularisation scheme the imaginary part of the effective Lagrangian (\ref{YMeffecLagrangian}) in the background chromomagnetic field   vanishes \cite{Savvidy:1977,Savvidy:2019grj}:
\beqa\label{imaginarypart}
\CI m~ \CL^{(1)}_{YM} (H)
 &=& -  {g H \over 4 \pi^2} \int^{\infty}_{0} {ds \over s^2}      
  \sin( \mu^2 s)  \sin (g H s)     + {g^2 H^2 \over 4 \pi^2} \int^{\infty}_{0} {ds \over s} \sin( \mu^2 s)    \nn\\
&&\nn  \\
  &=& -  {g H \over 4 \pi^2}   {\pi \over 2 } g H      + {g^2 H^2 \over 4 \pi^2}  {\pi \over 2 }    
= -  {g^2 H^2  \over 8 \pi}      + {g^2 H^2 \over 8 \pi}   =0.   
\eeqa
The presence/absence of the imaginary parts in the Yang-Mills effective action has been a source of controversy and therefore requires additional physical arguments and proofs to confirm the above conclusion.

In the case of a pure chromomagnetic field  the spectrum of the operator $H(1)$ has the following form  \cite{Nielsen:1978rm,Skalozub:1978fy,Ambjorn:1978ff}:
\be\label{negmode}
k^{2}_0 =  k^2_{3} + (2n+1 \pm 2) g H, ~~~n=0,1,2,...,
\ee
and due to the spin-interaction term $2g\hat{G}$ there is a negative mode $k^{2}_0 =  k^2_{\vert\vert} - g H$ when  $n=0$  ( $k^2_{\vert\vert} \leq g H $). This mode can induce the  {\it imaginary term} of the effective Lagrangian of  the following form (formula (2.36) in \cite{Nielsen:1978rm}):
\be\label{NOcontribution}
2 \CI m ~\CL^{(1)}(H) = \frac{e H}{4\pi^2} \CI m \int^{\infty}_{-\infty} d k_3 \sqrt{k^2_{3} - g H -i \epsilon} = - \frac{g^2 H^2}{4 \pi}. 
\ee
Since the vacuum-persistence probability is given by the formula (expression (5.31) in  \cite{Schwinger:1951nm} ):
\be\label{probabi1}
|\langle 0\vert 0 \rangle|^2=  \left| \exp \left\{ i \Gamma(H) \right\} \right|^{2} = \exp \left\{ -2~ \mathcal{I}m ~\Gamma(H) \right\},
\ee
the imaginary part of $\Gamma$ defines the decay rate of the vacuum and therefore imposes  a strong physical constraint 
\be 
2 ~\CI m \Gamma \geq 0
\ee
on any possible imaginary term  in the effective action because the probability must be less or equal to 1.

It follows that the negative imaginary part of the effective action (\ref{NOcontribution}) appearing due to the negative mode in the spectrum (\ref{negmode}) leads to an apparent conflict with the unitarity and causality of the theory  (\ref{probabi1}).  This inconsistency  points to the fact that the imaginary term is a result of the quadratic approximation and therefore requires analysis of the quantum fluctuations beyond the quadratic approximation \cite{Nielsen:1978nk, Ambjorn:1978ff, Ambjorn:1980ms,Flory:1983td,Savvidy:2019grj,Savvidy:2022jcr, Savvidy:2023kft, Savvidy:2023kmx}.  It appears that the self-interaction of the negative mode eliminates the imaginary term from the chromomagnetc effective action. 

Let us consider first a number of physical arguments and analytical results that lead to the conclusion that there are no  imaginary terms in the effective action in the case of  chromomagnetic gauge fields (\ref{YMeffecLagrangian}).  First of all, the magnetic field does no work  and therefore cannot separate  a pair of virtual charged particles  to the asymptotic states at infinity \cite{Savvidy:1977}, as it happens in the case of the electric field \cite{Sauter:1931zz, Heisenberg:1936nmg, Schwinger:1951nm}.  Secondly,  the probability (\ref{probabi}) that no actual pair creation occurs during the history of the system evolution leads to the inequality  
\be
2~ \mathcal{I}m ~\Gamma(H) \geq 0
\ee 
meaning  that any imaginary term in the effective action should be non-negative, otherwise it will break the unitarity and causality of the theory.   

Next let us consider the structure of the effective Lagrangians (\ref{YMeffLagr} ), (\ref{HeiEu1}) and (\ref{HeiEu2}).   The  "particle" Hamiltonians defining the matrix element $(x'(s) | x''(0) )$ in QCD (\ref{foperats}), (\ref{matrixelement}), (\ref{functequation}) and QED (\ref{qedpart}) contain the orbital interaction term $\Pi^2_{\mu} = (i \nabla_{\mu})^2$. The operator $L(s)$ in (\ref{lsoper}), (\ref{Lmatrix}) represents the contribution of the orbital interaction term  in the matrix element $(x'(s) | x''(0) )$ and appears in all the three effective Lagrangians (\ref{YMeffLagr}), (\ref{HeiEu1}), and (\ref{HeiEu2}) in the following form: 
\be\label{denomin}
\frac{e f_1 s \cdot e f_2 s }{\sinh(e f_1 s) \sin(e f_2 s)}.
\ee
The contribution of the spin-interaction term $ \frac{1}{2} e \sigma_{\mu\nu} F_{\mu\nu}$ in QED results into the expression  (\ref{HeiEu1})
\be\label{nomi1}
 \cosh(e f_1 s) \cos(e f_2 s),
\ee
and in QCD the spin interaction term  $2 g \hat{G}_{\mu\nu}$ results into the expression  (\ref{Nmatrix})
\be\label{nomi2}
\cos(2 g f_1 s) + \cosh(2 g f_2 s).
\ee
The singularities in the finite part of the complex plane  $\textcircled{s} $ can only be created by the functions in the denominator (\ref{denomin}) resulting from the orbital interaction term $\Pi^2_{\mu}$. The functions in the nominator (\ref{nomi1}) and (\ref{nomi2}) are from the spin-interaction terms and don't  create singularities in the finite part of the complex plane  $\textcircled{s} $.  The conclusion that can be derived from this consideration is that  the spin-interaction terms don't  contribute to the imaginary terms of the effective action for any background  field. 

To support this statement further one should consider not only quadratic fluctuations of the negative mode amplitude but also its nonlinear self-interaction \cite{Nielsen:1978nk, Ambjorn:1978ff, Ambjorn:1980ms,Flory:1983td,Savvidy:2019grj,Savvidy:2022jcr, Savvidy:2023kft, Savvidy:2023kmx}.  The quadratic approximation of the effective action,  that is, the one-loop approximation (\ref{oneloopaction}),  becomes inadequate in this circumstance.  The self-interaction of the negative mode was considered by Ambjorn, Nielsen, Olesen, Flory and other authors \cite{Nielsen:1978nk, Ambjorn:1978ff,  Ambjorn:1980ms, Flory:1983td, Kay:1983an, Parthasarathy:1983ck, Kim:2016xdn}, who came to the conclusion that self-interaction eliminates the imaginary term in the effective Lagrangian (see also Appendix D).  

The contribution of quadratic and nonlinear self-interaction terms of the negative mode to the effective action was considered  recently in \cite{Savvidy:2019grj,Savvidy:2022jcr, Savvidy:2023kft, Savvidy:2023kmx}.  The eigenfunction of a charged vector boson in a magnetic field that corresponds to the negative mode has the following form \cite{Nielsen:1978nk, Ambjorn:1978ff}:  
\be\label{negativemode}
e(x_0,x_1,x_2,x_3)=e^{- {1\over 2} g H (x_1 - k_2/g H)^2 + i (k_2  x_2  +k_{3} x_{3} - k_0 x_0)} .  
\ee
By introducing the dimensionless amplitude $a_{k_{2}}(x_3,x_0)$ of the gauge field in the subspace of the negative mode one can represent the amplitude in the following form\footnote{The  negative-mode amplitude has  the following form 
$W_1 = -i W_2 = W = \frac{1}{\sqrt{2}}(w_1 +i w_2)$, $W_0=W_3=0$,  where $w_1(x,y), w_2(x,y) $ are real and imaginary parts of the charged field $W_{\mu}= \frac{1}{\sqrt{2}}(A^1_{\mu} +i A^2_{\mu})$  \cite{Nielsen:1978nk, Ambjorn:1978ff}.}:
\be\label{negmodampli}
W(x_0,x_1,x_2,x_{3}) =  \frac{1}{2^{1/4}} \int\frac{d k_2}{2\pi}    e^{- {1\over 2} g H (x_1 - k_2/g H)^2 + i k_2  x_2   } a_{k_{2}}(x_{3},x_0).
\ee
The part of the Yang-Mills classical action representing the negative mode that includes now the quadratic and the self-interaction terms is \cite{Nielsen:1978nk, Ambjorn:1978ff}\footnote{The formulas (21),(23) in \cite{Nielsen:1978nk} and (3.4),(3.6) in \cite{Ambjorn:1978ff}. } 
\beqa\label{negmodact}
&S_{negative~mode} =  \sqrt{\frac{2 \pi}{g H}}  \int \frac{ d k_2}{2\pi}  \int d x_{3} d x_0  \{ - \vert \partial_{\mu} a_{k_{2}}  \vert^2 + g H \vert a_{k_{2}}  \vert^2 \} - \nn\\
&- \frac{g^2}{2}  \sqrt{\frac{2 \pi}{g H}}   \int \frac{d k_2 d p  d q } {(2\pi)^3}    e^{ -\frac{ p^2 +q^2} {2 gH }}   
  \int  d x_{3} d x_0  ~a^*_{k_2 +p}  a^*_{k_2 +q} a_{k_2} a_{k_2 +p +q}. 
\eeqa 
The first term, quadratic  in amplitude, $a_{k_{2}}$, represents the negative mode (\ref{negmode}) with its negative frequency  $ g H \vert a_{k_{2}}  \vert^2 $, while the second term represents its self-interaction term. Now the functional integral over the amplitude $a_{k_{2}}$ gets a dominant  contribution from the positive definite quartic-interaction potential $\propto g^2   a^4_{k_{2}} $ in (\ref{negmodact}) that provides convergence of the functional integral and eliminates any remnants of the imaginary term. Now the question is how the self-interaction term will change the real part of the effective Lagrangian  (\ref{YMeffLagr}), (\ref{YMeffecLagrangian}) and (\ref{savvacgg}). This question appears because initially the contribution of the negative mode to the effective Lagrangian was considered in the quadratic approximation (\ref{vectorhamilt1}), (\ref{savvacgg}) and (\ref{NOcontribution}).

Thus the problem reduces to the exact evaluation of the functional integral over the quartic interaction of the negative-mode amplitude.   Miraculously, the problem can  be solved  due to the conformal invariance of the classical action (\ref{negmodact}) and the functional integral over the negative-mode amplitude can be evaluated exactly \cite{Savvidy:2019grj, Savvidy:2022jcr, Savvidy:2023kft, Savvidy:2023kmx}. The functional integration can be performed by passing to the dimensionless variables $k_{\mu} \rightarrow k_{\mu}/ \sqrt{gH}$, $x_{\mu} \rightarrow x_{\mu}  \sqrt{gH}$.  For the negative-mode amplitude (\ref{negmodampli}) one can obtain  
\be\label{unstablemode}
W~ \Big({ \mu^2  \over g H } \Big)^{1/2}   =  \int {d k_2 \over 2 \pi} e^{-{1\over 2}(x_1 + k_2 )^2 + i k_2 x_2}  a_{k_{2}}(x_0,x_3),
\ee
while the action for the unstable mode amplitude  (\ref{negmodact}) will take the following form: 
\be\label{unstablelagrangian}
S_{negative\ mode}  =  \int {d k_{2} \over 2 \pi } d x_0 d x_3 \Big(- \vert \partial_{\mu} a_{k_{2}} \vert^2 +  \vert  a_{k_{2}} \vert^2 
- {1\over 2} g^2  \int {d p d q \over (2 \pi)^2 }  e^{-{p^2 +q^2\over 2}} a^-_{k_2 +p} a^-_{k_2 +q} a_{k_2} a_{k_2 +p+q}\Big).
\ee
In this representation  the dependence on the chromomagnetic field completely factorises from the action (\ref{unstablelagrangian}) and appears only in front of the negative-mode amplitude (\ref{unstablemode}) as $({\mu^2 \over gH})^{1/2}$. Therefore  the contribution of the negative-mode to the effective Lagrangian is only through the integration measure $ \prod_{k_2}  \CD a_{ k_{2} }  \simeq  \prod_{k_2} ({\mu^2 \over gH})^{1/2}$ and its degeneracy\footnote{ The quartic integral that remains in the exponent of the functional integral is a field-independent expression (\ref{unstablelagrangian}) and can be absorbed into the irrelevant integration constant $\CN$. From the expansion of the functional integral over the coupling constant $g^2$ it also follows that this contribution of the negative mode is a sum of all loop diagrams  with the negative mode propagating in the loops.} $({gH \over 2\pi})^2$: 
\be
Z_{negative~mode}   =  \CN \Big({ \mu^2  \over g H } \Big)^{{1\over 2}  ({gH \over 2\pi})^2 } = \CN e^{- {g^2 H^2 \over 8 \pi^2}  \log {g H \over \mu^2}  ~}.
\ee  
This contribution to the effective Lagrangian is a real function of the chromomagnetic field 
\be\label{negmodecon}
\CL_{negative~mode}(H)= - {g^2 H^2 \over 8 \pi^2}  \log {g H \over \mu^2},
\ee 
and together with the contribution of the positive modes in (\ref{negmode}) 
\be\label{positmodecon}
\CL_{positive~modes}(H) =- {5 g^2 H^2 \over 48 \pi^2}  \log {g H \over \mu^2} 
\ee
the effective Lagrangian takes the form   (\ref{savvacgg}) without imaginary term. Thus the outcome of the functional integration, the sum of the (\ref{negmodecon}) and (\ref{positmodecon}), confirms that the contribution of the negative mode does not change the real part of the effective Lagrangian  (\ref{savvacgg}) \cite{Savvidy:1977as}.  The underlying physical reason lies in the fact that the functional integral over the negative mode amplitude measures the entropy, the Landau degeneracy of the negative mode.
 
A further support of the conjecture that {\it the spin-interaction terms don't  contribute to the imaginary terms of the effective Lagrangian in sourceless background  field} is due to the  Leutwyler  consideration of the vacuum polarisation by constant self-dual field configurations \cite{Leutwyler:1980ev, Leutwyler:1980ma,Minkowski:1981ma}. The corresponding spectrum of $H(1)$ {\it has only positive modes} and infinite many zero modes, so called chromons, and therefore demonstrates the absence of imaginary terms in the effective Lagrangian \cite{Leutwyler:1980ev,Leutwyler:1980ma,Minkowski:1981ma}.  The exact contribution of zero-mode chromons to the effective action  was recently evaluated in \cite{Savvidy:2023kft,Savvidy:2023kmx}.

Let us now turn to the case of the chromoelectric field $\mathcal{G} = 0 $, $ \mathcal{F} < 0$:
\begin{equation}
f_{1}=0, \quad f_{2}=\sqrt{-\CF} \equiv E,
\end{equation}
 so that we will have
\begin{equation}\label{chroelec}
\mathcal{L}_{YM}^{(1)}(E)=\frac{1}{8 \pi^{2}} \int_{s_{0}}^{\infty} \frac{d s}{s^{3}} \frac{g E s}{\sin(g E s)}-
\frac{1}{4 \pi} \int_{s_{0}}^{\infty} \frac{d s}{s^{3}} g E s \sin(g E s).
\end{equation}
The integral over the proper time has singularities at
\be
s=s_n=\pi n / g E,~~~~ n=1, 2,...
\ee
and the Lagrangian (\ref{chroelec}) will develop a {\it positive imaginary contribution} to $\CL_{YM}$ \cite{Savvidy:1977,Savvidy:2019grj}:
\begin{equation}\label{imagin}
2 \CI m ~ \mathcal{L}_{YM}^{(1)}(E) = \frac{(gE)^{2}}{4\pi^{3}} \sum_{n=1}^{\infty} \frac{(-1)^{n+1}}{n^{2}} = \frac{g^{2}E^{2}}{48\pi}.
\end{equation}
This is the probability, per unit time and per unit volume, that a pair is created by the constant chromoelectric field. The probability of all the processes with the conservation of the vacuum state is defined by the quantity 
\be
\left| \exp \left\{ i \Gamma(E) \right\} \right|^{2} = \exp \left\{ -2\mathcal{I}m~ \Gamma(E) \right\} =  \exp \left\{-  \frac{g^{2}E^{2}}{48\pi} V T \right\} ,
\ee
and (\ref{imagin}) provides a  decay rate of the constant chromoelectic field.  

In the next two sections we will calculate the effective Lagrangian for the chromomagnetic flux tube solutions considered in the previous sections, in particular, we will consider the "polynomial" and "hyperbolic" solutions. It seems that the same technic can also be used for the evaluation of the effective Lagrangian for the general solution (\ref{magneticsheetsolution1}).

\section{\it Effective action for polynomial flux tube solution}

Here we will compute the effective Lagrangian for the polynomial solution (\ref{polsol}), (\ref{magsheet}) in the limit  $a \rightarrow 0 $  when the chromomagnetic field spreads over all the 3D-space and by considering $b \rightarrow \infty$  while keeping the product  $a b$ fixed  in order to obtain a finite energy-density solution $ \epsilon = {  a^2 b^2  \over 2 g^2}$ in the whole 3d-space.  This solution of the Yang-Mills equation (\ref{choansatzint}) 
\be\label{choansatz1}
A^{a}_{\mu} =  B_{\mu} n^{a}  +
{1\over g} \varepsilon^{abc} n^{b} \partial_{\mu}n^{c}
\ee
has the Abelian part
\be
 B_{\mu} =\{ 0, B_1,0,0 \},~~~~B_1 = -H y,~~~F_{12}(B)  = H,
\ee
and the part associated with the unit colour vector $n^a(x,y)$ of the form
\be\label{fluxsheet12}
n^a = \{ {\sqrt{1 - (ax)^2}}\,\cos (by),{\sqrt{1 - (ax)^2}}\,\sin (by), ax\},~~~~ g G_{12} = g H - a b.
\ee
It is convenient to calculate first the spectrum of the Faddeev-Popov  Hamiltonian $H^{ab}_{0}$ (\ref{vectorhamilt1}),
\beqa
H^{ab}_{0} = \nabla^{ac}_{\mu}(A) \nabla^{cb}_{\mu}(A),~~~~~~~~
\eeqa
by representing the covariant derivative $\nabla^{ab}_{\mu}(A)=
  \delta^{ab} \partial_{\mu}   - g \varepsilon^{acb} A^{c}_{\mu}$ in the following form:
\beqa
\nabla^{ab}_{\mu}(A)= \delta^{ab} \partial_{\mu}- g B_{\mu} \hat{n}^{ab} +
n^{a} \partial_{\mu} n^{b}  -  n^{b} \partial_{\mu} n^{a} =
\nabla^{ab}_{\mu}(B) + {\cal A}^{ab}_{\mu},
\eeqa
where 
\beqa
\nabla^{ab}_{\mu}(B) = \delta^{ab} \partial_{\mu}- g B_{\mu} \hat{n}^{ab},~~~~~~~~~
{\cal A}^{ab}_{\mu}  = n^{a} \partial_{\mu} n^{b}  -  n^{b} \partial_{\mu} n^{a}.
\eeqa
The  operator $H^{ab}_{0} $  will takes the following form:
\beqa
H^{ab}_{0} = \nabla^{ac}_{\mu}(B) \nabla^{cb}_{\mu}(B) -g B_{\mu} \hat{n}^{ac} {\cal A}^{cb}_{\mu}
-{\cal A}^{ac}_{\mu} g B_{\mu} \hat{n}^{cb} +(\partial_{\mu}  {\cal A}^{ab}_{\mu}) +
{\cal A}^{ac}_{\mu} {\cal A}^{cb}_{\mu} +
  2 {\cal A}^{ab}_{\mu} \partial_{\mu},
\eeqa
where 
\beqa
(\partial_{\mu}  {\cal A}^{ab}_{\mu}) &=&n^{a} \partial^{2}_{\mu} n^{b} -
n^{b} \partial^{2}_{\mu} n^{a},~\nn\\
{\cal A}^{ac}_{\mu} {\cal A}^{cb}_{\mu} &=&-
n^{a} n^{b} \partial _{\mu} n^{c} \partial _{\mu} n^{c} -
\partial _{\mu} n^{a} \partial _{\mu} n^{b}, \nn\\
2{\cal A}^{ab}_{\mu} \partial_{\mu} &=&
2(n^{a} \partial_{\mu} n^{b}  -  n^{b} \partial_{\mu} n^{a} )\partial_{\mu}
\eeqa
and is a sum of two terms $H^{ab}_{0} = \tilde{H}^{ab}_{0}  + \tilde{\tilde{H}}^{ab}_{0}$:
\beqa
\tilde{H}^{ab}_{0} &=& (\delta^{ac}  \partial_{\mu}   - g B_{\mu} \hat{n}^{ac})
(\delta^{cb}  \partial_{\mu}   - g B_{\mu} \hat{n}^{cb})-g B_{\mu} \hat{n}^{ac} {\cal A}^{cb}_{\mu}
-{\cal A}^{ac}_{\mu} g B_{\mu} \hat{n}^{cb}, \\
\tilde{\tilde{H}}^{ab}_{0} &=& n^{a} \partial^{2}_{\mu} n^{b} -
n^{b} \partial^{2}_{\mu} n^{a} -
n^{a} n^{b} \partial _{\mu} n^{c} \partial _{\mu} n^{c} -
\partial _{\mu} n^{a} \partial _{\mu} n^{b}
+ 2(n^{a} \partial_{\mu} n^{b}  -  n^{b} \partial_{\mu} n^{a} )\partial_{\mu} .\nn
\eeqa
In the case of the polynomial solution (\ref{fluxsheet12}) the components of the gauge field are 
\beqa
&{\cal A}^{ab}_{\mu} =\{ 0, {\cal A}^{ab}_1 ,{\cal A}^{ab}_2,0 \},~~~{\cal A}^{ab}_1 = n^{a} \partial_{x} n^{b}  -  n^{b} \partial_{x} n^{a},~~~
 {\cal A}^{ab}_2 = n^{a} \partial_{y} n^{b}  -  n^{b} \partial_{y} n^{a} \nn\\
& F_{12}=H,~~~~~S _{12} = \varepsilon^{abc} n^a \partial_{x} n^{b} \partial_{y} n^{c}=ab,~~~~ G_{12}= {g H - ab \over g} \nn\\
&\epsilon ={1\over 4}G^{a}_{ij} ~G^{a}_{ij}=   {(g H - ab)^2\over 2 g^2},
\eeqa
and the operator $ H^{ab}_{0}$ will take the following form:
\beqa
\tilde{H}^{ab}_{0} &=&  \delta^{ab}  \partial^{2}_{0} - \nabla^{ac}_{1}(B) \nabla^{cb}_{1}(B)
- \delta^{ab}  \partial^{2}_{2} - \delta^{ab}  \partial^{2}_{3} -g B_{1} \hat{n}^{ac} {\cal A}^{cb}_{1}
-{\cal A}^{ac}_{1} g B_{1} \hat{n}^{cb}, \\
\tilde{\tilde{H}}^{ab}_{0} &=&-  n^{a} \partial^{2}_{i} n^{b} +
n^{b} \partial^{2}_{i} n^{a} +
n^{a} n^{b} \partial _{i} n^{c} \partial _{i} n^{c} +
\partial _{i} n^{a} \partial _{i} n^{b}
- 2(n^{a} \partial_{i} n^{b}  -  n^{b} \partial_{i} n^{a} )\partial_{i}. \nn
\eeqa
The eigenvalues equation
\beqa
H^{ab}_{0} \Psi^b = \Lambda \Psi^a
\eeqa
can be projected into the orthonormal frame  (see Appendix B):
\be\label{orthonormalframe}
n^a,~~~e^a_{1}  ={1\over a} \sqrt{1 - (ax)^2} ~\partial_{x} n^a,~~~
e^a_{2} = \frac{1}{b \sqrt{1 - (ax)^2} } \partial_{y}  n^a~.
\ee
The expansion of the wave function is
\beqa
\Psi^b = \xi ~n^b  +  \eta ~e^{b}_1  +\varsigma ~ e^{b}_2,
\eeqa
where its components are $\xi(t,x,y,z), ~\eta(t,x,y,z),~  \varsigma(t,x,y,z)$. 
Calculating the action of the $H^{ab}_0$ on the first component $~\xi ~n^b$ we will get 
\beqa
H^{ab}_{0} ~\xi ~n^b=  n^{a} \partial^{2}_{\mu} ~\xi,
\eeqa
where we used the matrix elements given in the Appendix C (see (\ref{matrixelements1})). The action of $H_0$ on the component $\eta ~e^{b}_1$ is
 \beqa
H^{ab}_{0} ~\eta ~e^{b}_1 =
e^{a}_1 \Big( \partial^{2}_{\mu}  ~\eta  + a^2 b^2 x^2 \eta +  g^2 H^2 y^2 \eta  \Big) 
+ e^{a}_2 \Big(2  g H y~\eta^{'}_{x} +2 a b x ~\eta^{'}_{y}  \Big)   
\eeqa
and on the $~\varsigma ~e^{b}_2$ is
\beqa
H^{ab}_{0} ~\varsigma ~e^{b}_2=  e^a_2 \Big(\partial^{2}_{\mu}  ~\zeta  +  a^2 b^2 x^2 \zeta + g^2 H^2 y^2 \zeta \Big)-
e^{a}_1 \Big( 2 g H y~\zeta^{'}_{x}+ 2 a b x ~\zeta^{'}_{y} \Big). 
\eeqa
The projection of the equation  $H^{ab}_0 \Psi^b$  into the orthonormal frame (\ref{orthonormalframe}) gives
\beqa
&&\partial^{2}_{\mu} ~\xi , \nonumber  \\
&&\partial^{2}_{\mu} ~\eta ~ - ~2 g H y ~\zeta^{'}_{x}~ -~ 2 a b ~ x   ~\varsigma^{'}_{y}
~+ ~g^2 H^2 y^2 ~\eta ~+~
 a^2 b^2~ x^2 ~\eta~, \nonumber  \\
&&\partial^{2}_{\mu} ~\varsigma ~+~  2 g H y ~\eta^{'}_{x}~
+ ~2 a b~ x    ~\eta^{'}_{y}~
+~g^2 H^2 y^2~\varsigma ~+~
a^2 b^2~x^2  ~\varsigma   .
\eeqa
For  the charge component
\beqa
\phi  = \eta + i \varsigma
\eeqa
we will obtain the following equation:
\beqa\label{ghostespecquation}
\partial^{2}_{\mu} ~\phi~ + ~2 i g H y ~\phi^{'}_{x}~
+~ 2 i a b  x   ~\phi^{'}_{y} 
~+ ~g^2 H^2 y^2 ~\phi ~+~
a^2 b^2 x^2 ~\phi= 0,
\eeqa
or in an equivalent form as
\beqa
\partial^{2}_{0} ~\phi~ -\partial^{2}_{z} ~\phi
+(i\partial_{y}   ~+~ a b x  )^2 ~\phi~ + (i \partial_{x} ~+~  g H y )^{2}~\phi = \Lambda \phi.
\eeqa
By searching the solution of the equation in the following form:
\be
\phi(t,x,y,z) = \int  \frac{d k_0}{2 \pi} \frac{d k_3}{2\pi} e^{i k_0 t - i k_3 z} \psi(k_0,k_3,x,y),
\ee
we will obtain 
\be
\Big( -k^2_0 +k^2_3 + (i\partial_{y}   ~+~ a b x  )^2 ~ + (i \partial_{x} ~+~  g H y )^{2}~\Big) \psi  = \Lambda \phi.
\ee
Two operators naturally appearing in the above equation  $(v_x,v_y)$,
\beqa\label{velocityoper}
&v_x= i \partial_{x} ~+~  g H y,~~~~~v_y=i\partial_{y}   ~+~ a b x, 
\eeqa
have the following  commutation relation:
\beqa 
 ~[v_y, v_x] = i (gH-a b)= i g G_{12}.
\eeqa
These operators can be identified with the standard Heisenberg operators $(P,Q)$:
\be
v_x = P ,~~~v_y= Q (gH-a b),~~~[Q,P]=i,
\ee
and  the spectrum of the Faddeev-Popov ghost $H_0$ operator will coincide with the spectrum of the harmonic oscillator of the frequency $\omega^2 = (gH -a b)^2$:
\be
\Big( - k^2_0 + k^2_3 + P^2 + Q^2(gH -a b)^2 \Big) \psi= \Lambda \phi.
\ee
Thus the  spectrum  of the ghost Hamiltonian has the following form:
\be\label{ghostmods}
\Lambda =- k^2_0 + k^2_3 + (2n +1) \vert gH - ab \vert,~~~~~~n=0,1,2,....
\ee
The eigenfunctions of the Hamiltonian $H^{ab}_{\mu\nu} $
\beqa\label{vectoreigeneq}
\Big( g_{\mu\nu} H^{ab}_0   +2 g G^{ab}_{\mu\nu}  \Big) \Psi^b_{\nu}   =  \Lambda \Psi^a_{\mu}
\eeqa
can also be expanded into the orthonormal frame
\be
 \Psi^b_{\nu} =  \xi_{\nu}~n^b    +   \eta_{\nu}~e^b_{1}    +  \zeta_{\nu} ~e^b_{2} .
\ee
The spin interaction term can be represented in the following form:
\beqa
2 g G^{ab}_{\mu\nu} \Psi^b_{\nu}  = 2 g G_{\mu\nu} \hat{n}^{ab}  \Psi^b_{\nu} = 
2 g G_{\mu\nu} \hat{n}^{ab} ( \xi_{\nu}~n^b    +   \eta_{\nu}~e^b_{1}    +  \zeta_{\nu} ~e^b_{2}  )= 
2 g G_{\mu\nu} \Big (-  \eta_{\nu}~  e_{2}^{a}     + \zeta_{\nu} ~  e_{1}^{a}   \Big), \nn
\eeqa
where the following  relations   were used:
\be
\hat{n}^{ab} n^b=0,~~~\hat{n}^{ab} e_1^{b} = -e^a_2  ,~~~\hat{n}^{ab} e^{b}_2 = e^{a}_1 .
\ee
 The equation  (\ref{vectoreigeneq}) will take the following form:
\beqa
&& \Big( g_{\mu\nu} H^{ab}_0   +2 g G^{ab}_{\mu\nu}  \Big) (~n^b \xi_{\nu}   +  ~e^b_{1}  \eta_{\nu}   +  ~e^b_{2} \zeta_{\nu} )= \nn\\
&&= H^{ab}_0 n^b  \xi_{\mu} +  H^{ab}_0 e^b_1  \eta_{\mu} +  H^{ab}_0 e^b_2 \zeta_{\mu}  -2 g G_{\mu\nu} \eta_{\nu} ~ e_{2}^{a}     + 2 g G_{\mu\nu}  \zeta_{\nu}~  e_{1}^{a} .
\eeqa
Projecting the  equation into the orthonormal frame (\ref{orthonormalframe}) one can get
\beqa
&&n^a \Big( g_{\mu\nu} H^{ab}_0   +2 g G^{ab}_{\mu\nu}  \Big) (~n^b \xi_{\nu}   +  ~e^b_{1}  \eta_{\nu}   +  ~e^b_{2} \zeta_{\nu} )= \partial^2_{\mu} \xi_{\nu} ,\nn\\
&&e^a_1 \Big( g_{\mu\nu} H^{ab}_0   +2 g G^{ab}_{\mu\nu}  \Big) (~n^b \xi_{\nu}   +  ~e^b_{1}  \eta_{\nu}   +  ~e^b_{2} \zeta_{\nu} )=e^a_1 H^{ab}_0 e^b_{1} ~ \eta_{\mu}  + e^a_1 H^{ab}_0 e^b_{2}~  \zeta_{\mu} + 2 g G_{\mu\nu} \zeta_{\nu},\nn\\
&&e^a_2 \Big( g_{\mu\nu} H^{ab}_0   +2 g G^{ab}_{\mu\nu}  \Big) (~n^b \xi_{\nu}   +  ~e^b_{1}  \eta_{\nu}   +  ~e^b_{2} \zeta_{\nu} )=e^a_2 H^{ab}_0 e^b_{1} ~ \eta_{\mu}  + e^a_2 H^{ab}_0 e^b_{2}~  \zeta_{\mu} - 2 g G_{\mu\nu} \eta_{\nu}\nn
\eeqa
and then using the matrix elements (\ref{matrixelements1})   obtain the system of equations
\beqa
&&  \partial^2_{\lambda} \xi_{\mu}=\Lambda \xi_{\mu}, \nn\\
&& \partial^2_{\lambda}~ \eta_{\mu}  - 2 g H y ~  \partial_x \zeta_{\mu} - 2  a b x~  \partial_y  \zeta_{\mu}+(g^2 H^2 y^2 + a^2 b^2 x^2)~ \eta_{\mu}  + 2 g G_{\mu\nu} \zeta_{\nu} =\Lambda \eta_{\mu},\nn\\
&&\partial^2_{\lambda}~ \zeta_{\mu}  +2 g H y ~  \partial_x \eta_{\mu} + 2  a b x~  \partial_y  \eta_{\mu}+(g^2 H^2 y^2 + a^2 b^2 x^2)~ \zeta_{\mu}  - 2 g G_{\mu\nu} \eta_{\nu} =\Lambda \zeta_{\mu}.
\eeqa
Introducing the charged field $\Phi_{\mu} = \eta_{\mu} +i \zeta_{\mu}$ we will get 
\beqa
 && \partial^2_{\lambda} \xi_{\mu}=0 \\
&& \partial^2_{\lambda}~ \Phi_{\mu} + 2i  g H y ~  \partial_x \Phi_{\mu} + 2 i a b x~  \partial_y  \Phi_{\mu}+(g^2 H^2 y^2 + a^2 b^2 x^2)~ \Phi_{\mu}  - 2i g G_{\mu\nu}\Phi_{\nu} =\Lambda \Phi_{\mu}. \nn
\eeqa
The neutral component of the vector boson in the first equation has a trivial spectrum and the second equation can be represented in the following equivalent  form:
\beqa
 \partial^2_{0}~ \Phi_{\mu} + (i \partial_x + g H y)^2 ~  \Phi_{\mu} + (i \partial_y +  a b~x)^2   \Phi_{\mu}  - 2i g G_{\mu\nu}\Phi_{\nu} =\Lambda \Phi_{\mu}.
\eeqa
Using the operators $(v_x,v_y)$ (\ref{velocityoper}) and their Heisenberg realisation we obtain the system of equations 
\beqa\label{gaugebosoneigen}
 ( -k^2_0 +k^2_3 + P^2 + Q^2(gH -a b)^2) \Phi_1 - 2 g G_{12} \Phi_2=\Lambda \Phi_{1} , \nn\\
 (-k^2_0 +k^2_3 + P^2 + Q^2(gH -a b)^2) \Phi_2 + 2 g G_{12} \Phi_1=\Lambda \Phi_{2},
  \eeqa
with the  spectrum of the following form:
\be\label{gaugebosoneigen}
\Lambda   = -k^2_0 + k^2_3 +(2n +1 \pm 2) \vert gH - ab \vert , ~~~~n =0,1,2,....
\ee
The contribution of the longitudinal gauge boson modes
\beqa\label{longitudinalmodes}
  (-k^2_0 +k^2_3 + P^2 + Q^2(gH -a b)^2) \Phi_0 = \Lambda \Phi_0,\nn\\
   (-k^2_0 +k^2_3 + P^2 + Q^2(gH -a b)^2) \Phi_3 =\Lambda \Phi_3
 \eeqa
is cancelled in (\ref{oneloopaction})  by the contribution of the ghost modes (\ref{ghostmods}). 
The problem reduces to the calculation of the gauge boson determinant with the eigenvalues (\ref{gaugebosoneigen}).  Here again there is a negative mode $k^{2}_{0} =  k^2_3 - \vert gH - ab \vert$, and in order to calculate its contribution to the effective Lagrangian one should take into account the non-linear self-interaction  of the  negative mode as it was described in the seventh section (\ref{negmodecon}).

Above we were calculating the spectrum of the operators $H^{ab}_{\mu\nu} $ and $H^{ab}_{0} $ in  the case of  parallel vectors $\vec{H}$ and $\vec{a} \times \vec{b}$. In order to investigate polarisation effects and the effective Lagrangian in the case of a general orientation of these vectors we will consider the Abelian field of the following form:
\be
 B_{\mu} =\{ 0, B_1,0,0 \},~~~~B_1 = -H y +F z,~~~F_{12}(B)  = H, ~~~~F_{31}(B)  = F,
\ee
and the part associated with the unit colour vector $n^a(x,y)$ of the form
\be\label{fluxsheet13}
n^a = \{ {\sqrt{1 - (ax)^2}}\,\cos (by),{\sqrt{1 - (ax)^2}}\,\sin (by), ax\},~~~~ g G_{12} = g H - a b, ~~~~g G_{31} = g F,
\ee
so that the vectors $\vec{H}$ and $\vec{a} \times \vec{b}$ are under a nonzero angle:
\be
(g \vec{H} - \vec{a} \times \vec{b})^2 = g^2 H^2 - 2 \vert g \vec{H}\vert  \vert\vec{a} \times \vec{b} \vert \cos\gamma + \vert\vec{a} \times \vec{b} \vert^2, ~~~~   \cos \gamma = \frac{H}{\sqrt{H^2 +F^2}}.
\ee
In this case the following generalisation of the equation (\ref{ghostespecquation}) takes place:
\beqa\label{ghostespecquation1}
\partial^{2}_{\mu} ~\phi~ + ~2 i g (H y-F z) ~\phi^{'}_{x}~
+~ 2 i a b  x   ~\phi^{'}_{y} 
~+ ~g^2 (H  y - F z)^2 ~\phi ~+~
a^2 b^2 x^2 ~\phi= \Lambda \phi,
\eeqa
and it can be represented in the following form:
 \beqa\label{ghostespecquation2}
\partial^{2}_{0} ~\phi~ -\partial^{2}_{z} ~\phi
+(i\partial_{y}   ~+~ a b x  )^2 ~\phi~ + (i \partial_{x} ~+~  g (H y - F z) )^{2}~\phi = \Lambda \phi.
\eeqa
 Operators naturally appearing in the above equation  are $(v_x,v_y, v_z)$:
\beqa\label{velocityoper1}
&v_x= i \partial_{x} ~+~  g (H y - F z),~~~~~v_y=i\partial_{y}   ~+~ a b x, ~~~~~~~v_z = i \partial_{z},
\eeqa
and have the following  commutation relations:
\beqa 
 ~[v_y, v_x] = i (gH-a b)= i g G_{12}, ~~~~~~[v_x, v_z] = i g F= i g G_{31}, ~~~~~[v_y, v_z] =  0.
\eeqa
In order to exclude the $x$ variable from the equation we will represent the wave function in the form 
\be
\phi = e^{i a b x y }  \chi
\ee
so that
\be
v_x = e^{i a b x y }  ( i \partial_{x} ~+~   (gH- ab)  y - g F z ) \chi, ~~~v_y= e^{i a b x y }   i \partial_{y}\chi~~~, v_z= e^{i a b x y }   i \partial_{z}\chi,
\ee
and the equation transforms to the following form:
\beqa\label{ghostespecquation3}
\partial^{2}_{0} ~\chi~ -\partial^{2}_{z} ~\chi
-\partial^2_{y}  ~\chi~ + (i \partial_{x} ~+~   (gH -ab) y - g F z) )^{2}~\chi = \Lambda \chi.
\eeqa
We introduce  the new variables 
\be
u = (g H-ab) y - g F z,~~~w= \frac{g F y + (g H-ab) z}{\sqrt{ (gH -a b)^2 + g^2 F^2}}, 
\ee
and the equation takes the following form:
\beqa\label{ghostespecquation4}
\partial^{2}_{0} ~\chi~ -\partial^2_{w} ~\chi  -((g H-ab)^2 + g^2 F^2)\partial^{2}_{u}  ~\chi~ + (i \partial_{x} ~+~  u )^{2}~\chi = \Lambda \chi.
\eeqa
The new operators can be identified with the standard Heisenberg operators $(P,Q)$ 
\be
P=i \partial_{x} ~+~  u , ~~~ Q=   i \partial_{u},~~~[Q,P]=i,
\ee
so that
\beqa\label{ghostespecquation4}
\partial^{2}_{0} ~\chi
-\partial^2_{w}  ~\chi~ + P^{2}~\chi ~ +\omega^2 Q^2 ~\chi= \Lambda \chi ,
\eeqa
where $\omega^2 = (gH -a b)^2 + g^2 F^2=  ( g\vec{H} - \vec{a} \times \vec{b})^2$ is the frequency of the harmonic oscillator, and  the spectrum of the Faddeev-Popov ghost operator $H_0$   can be represented in the following form:
\beqa\label{ghostspectrum}
\Lambda = -k^2_0 +  k^2_3 + (2n +1) \vert g\vec{H} - \vec{a} \times \vec{b} \vert.
\eeqa
 It is now straightforward to calculate also the spectrum of the gauge field:
 \beqa\label{gaugespectrum}
\Lambda = -k^2_0 +   k^2_3 + (2n +1 \pm 2) \vert g\vec{H} - \vec{a} \times \vec{b} \vert .
\eeqa
The field dependence of the spectrum is Lorentz- and gauge-invariant: 
\be
2 g^2 \CF =  g^2 \frac{1}{2} G^a_{\mu\nu}G^a_{\mu\nu}= (gH -a b)^2 + g^2 F^2=  ( g\vec{H} - \vec{a} \times \vec{b})^2 ,
\ee
and it is therefore convenient to represent the spectrum in the explicitly invariant from:
\be
\Lambda = -k^2_0 +   k^2_3 + (2n +1 \pm 2) \vert 2 g^2 \CF \vert .  
\ee
The contribution of the $H_0$  to the effective Lagrangian can be calculated by using the spectrum (\ref{ghostspectrum}) of the $H_0$   (\ref{finoneloopeffact}):
\beqa\label{FPcont}
    i \int  \frac{ds}{s}  ~  \hat{\operatorname{tr}}~  e^{- i H_0 s}   &=& 2 i  {\it Deg}     \int \frac{ds}{s}  ~ \sum^{\infty}_{n=0} \frac{d k_0 d k_3}{(2\pi)^2}  e^{- i (-k^2_0 + k^2_3 + (2n+1) \vert 2 g^2 \CF \vert + \mu^2 ) s} \nn\\
 &=&- \frac{1}{8 \pi^2}  \int \frac{ds}{s^3}  e^{- \mu^2 s}~ \frac{\vert 2 g^2 \CF \vert s }{\sinh(\vert 2 g^2 \CF \vert  s) },
\eeqa
where the degeneracy of the modes is $ {\it Deg}  = \frac{g H}{2 \pi } \frac{4\pi }{a b}$. For the gauge boson contribution we will get 
\beqa
&-&  \frac{i}{2} \int  \frac{ds}{s}  ~  \hat{\operatorname{tr}}  \operatorname{tr}   e^{- i H(1) s}   = - i  {\it Deg}     \int \frac{ds}{s}  ~  \int \frac{d k_0 d k_3}{(2\pi)^2}  \sum^{\infty}_{n=0} e^{- i (-k^2_0 +k^2_3 + (2n+1\pm 2) \vert 2 g^2 \CF \vert  + \mu^2 ) s} - \nn\\
&-& 2i  {\it Deg}     \int \frac{ds}{s}  ~  \int \frac{d k_0 d k_3}{(2\pi)^2}  \sum^{\infty}_{n=0} e^{- i (-k^2_0 +k^2_3 + (2n+1 ) \vert 2 g^2 \CF \vert  + \mu^2 ) s} , 
\eeqa
where the second term is the contribution of the longitudinal modes (\ref{longitudinalmodes}), and will be canceled by the contribution (\ref{FPcont}) of the Faddeev-Popov determinant. Thus  for the effective Lagrangian we have the following expression:
\beqa\label{yangmillsav}
\CL_{YM}^{(1)} &=&   - i  {\it Deg}     \int \frac{ds}{s}  ~  \int \frac{d k_0 d k_3}{(2\pi)^2}  \sum^{\infty}_{n=0} e^{- i (-k^2_0 +k^2_3 + (2n+1\pm 2) \vert 2 g^2 \CF \vert  + \mu^2 ) s}\\
&=& \frac{1}{8 \pi^2}  \int_{s_0}^{\infty} \frac{ds}{s^3}  e^{- \mu^2 s}~ \frac{\vert 2 g^2 \CF \vert  s}{\sinh(\vert 2 g^2 \CF \vert  s) } +\frac{1}{4 \pi^2}  \int_{s_0}^{\infty}  \frac{ds}{s^3}  e^{-i \mu^2 s} \vert 2 g^2 \CF \vert s \sin(\vert 2 g^2 \CF \vert  s),\nn
\eeqa
where $  H^2 =  \CF = \frac{1}{4} G^a_{\mu\nu}G^a_{\mu\nu} > 0$ and $\mathcal{G} = 0 $. This Lagrangian coincides with the expression  (\ref{YMeffecLagrangian}) and (\ref{savvacgg}) if one use the invariant renormalisation that will be considered in the forthcoming section.

\section{\it Effective action for hyperbolic flux tube solution}

Let us consider the  "hyperbolic" solution  (\ref{hypersol}), which has infinite width in the $x$ direction unlike the polynimial solution  (\ref{polsol}) consider above  and  distributed over the whole 3D-space:
\be\label{hypersol}
n^a(x)= \{ {\cos(b y \cosh^2(a x) )\over  \cosh(a x)}, { \sin(b y \cosh^2(a x)) \over \cosh(a x)}, \tanh(a x)  \}.
\ee 
The corresponding orthonormal frame  has the following form :
\be\label{orthonormalframe1}
n^a,~~~e^a_{1}  ={\cosh(a x) \over a  }  ~\partial_{x} n^a - b y \sinh(2 a x)~ e^a_2,~~~
e^a_{2} = \frac{1}{b \cosh(a x)} \partial_{y}  n^a~.
\ee
The matrix elements $e^a_i \hat{n}^{ab} e^b_j$  are identical with the matrix elements of the polynomial solution (see Appendix C (\ref{matrixelements1})). The  wave function is 
$
\Psi^b = \xi ~n^b  +  \eta ~e^{b}_1  +\varsigma ~ e^{b}_2.
$
The action of the $H^{ab}_0$ on the first component $~\xi ~n^b$ is
\beqa
H^{ab}_{0} ~\xi ~n^b=  n^{a} \partial^{2}_{\mu} ~\xi,
\eeqa
where we used the matrix elements given in the Appendix C  (\ref{matrixelements1}). The action of $H_0$ on the component $\eta ~e^{b}_1$ is
 \beqa
H^{ab}_{0} ~\eta ~e^{b}_1 &=&
e^{a}_1 \Big( \partial^{2}_{\mu}  ~\eta  +  y^2 (a b -g H - a b \cosh(2 a  x ))^2   + \frac{1}{4}  b^2 \sinh^2(2 a x) \Big)\eta  +\nn\\
&&+ e^{a}_2 \Big(2 a^2 b y  \cosh(2 a x) \eta -2 y (a b -g H - a b \cosh(2 a  x ))~\eta^{'}_{x} + b   \sinh(2 a x) ~\eta^{'}_{y}  \Big)   \nn
\eeqa
 and on the $~\varsigma ~e^{b}_2$ is
\beqa
H^{ab}_{0} ~\varsigma ~e^{b}_2 &=& e^{a}_1 \Big( \partial^{2}_{\mu}  ~\zeta  + y^2 (a b -g H - a b \cosh(2 a  x ))^2   + \frac{1}{4}  b^2 \sinh^2(2 a x) \Big) \Big)\zeta  +\nn\\
&&+ e^{a}_2 \Big(-2 a^2 b y  \sinh(2 a x) \zeta + 2 y (a b -g H - a b \cosh(2 a  x ))~\zeta^{'}_{x} - b   \sinh(2 a x) ~\zeta^{'}_{y}  \Big).   \nn
\eeqa
By projecting the    $H^{ab}_0 \Psi^b$  into the orthonormal frame (\ref{orthonormalframe1}) and introducing  the charge component
$
\phi  = \eta + i \varsigma
$
we will obtain the following equation:
\beqa\label{ghostespecquation2}
&&\partial^{2}_{\mu} ~\phi~  + \Big( y^2 (a b -g H - a b \cosh(2 a  x ))^2   + \frac{1}{4}  b^2 \sinh^2(2 a x) \Big) ~\phi  +\\
 && +2 i a^2 b y \sinh(2 a x) \phi  +  i b \sinh(2 a x)\partial_{y} \phi - 2 i y (ab -g H -a b \cosh(2 a x)) \partial_{x}~\phi = 0,\nn
\eeqa
or in an equivalent form as
\beqa
\partial^{2}_{0} ~\phi~ -\partial^{2}_{z} ~\phi
+(i\partial_{y}   ~+~ \frac{b}{2}  \sinh(2 a x)  )^2 ~\phi~ + (i \partial_{x} -  y (a b - g H -ab \cosh 2 ax) ~)^{2}~\phi = \Lambda \phi.~~~
\eeqa
By searching the solution of the equation in the following form:
\be
\phi(t,x,y,z) = \int  \frac{d k_0}{2 \pi} \frac{d k_3}{2\pi} e^{i k_0 t - i k_3 z} \psi(k_0,k_3,x,y),
\ee
we will obtain 
\be
\Big( -k^2_0 +k^2_3 + (i\partial_{y}   ~+~ \frac{b}{2}  \sinh(2 a x) )^2~ + (i \partial_{x} +  y ( ab \cosh 2 ax - a b + g H ) ~)^{2}~\Big) \psi  = \Lambda \phi.
\ee
Two operators naturally appearing in the above equation  $(v_x,v_y)$,
\beqa\label{velocityoper1}
&v_x= i \partial_{x} ~+~  ( ab \cosh 2 ax - a b + g H ) y,~~~~~v_y=i\partial_{y}   ~+~ \frac{b}{2}  \sinh(2 a x) , 
\eeqa
have the following  commutation relation:
\beqa 
 ~[v_y, v_x] = i (gH-a b)= i g G_{12}.
\eeqa
These operators can be identified with the standard Heisenberg operators $(P,Q)$:
\be
v_x = P ,~~~v_y= Q (gH-a b),~~~[Q,P]=i,
\ee
and  the spectrum of the  ghost $H_0$ operator coincides with the spectrum of the harmonic oscillator of the frequency $\omega^2 = (gH -a b)^2$:
\be
\Big( - k^2_0 + k^2_3 + P^2 + Q^2(gH -a b)^2 \Big) \psi= \Lambda \phi.
\ee
Thus the  spectrum  of the ghost Hamiltonian has the following form:
\be\label{ghostmods1}
\Lambda =- k^2_0 + k^2_3 + (2n +1) \vert gH - ab \vert,~~~~~~n=0,1,2,....
\ee
In a similar way one can obtain the spectrum  of the Hamiltonian $H^{ab}_{\mu\nu} $
\beqa\label{vectoreigeneq1}
\Big( g_{\mu\nu} H^{ab}_0   +2 g G^{ab}_{\mu\nu}  \Big) \Psi^b_{\nu}   =  \Lambda \Psi^a_{\mu}
\eeqa
with the eigenvalues:
\be\label{gaugebosoneigen1}
\Lambda   = -k^2_0 + k^2_3 +(2n +1 \pm 2) \vert gH - ab \vert , ~~~~n =0,1,2,....
\ee
As far are this spectrum is identical with the spectrum of the polynomial solution, the effective Lagrangian coincides with the expressions  (\ref{yangmillsav}), (\ref{YMeffecLagrangian}) and (\ref{savvacgg}) obtained  for a constant gauge field (\ref{consfield}). 

Based on the universal form of the spectrum and of the effective Lagrangian that was obtained for these    field configurations one can conjecture that the effective Lagrangian for the general chromomagnetic flux solution (\ref{magneticsheetsolution1}) has a universal form (\ref{savvacgg}).   
The conclusion that can be drawn  from this result is that the Yang-Mills vacuum state is highly degenerate with the vacuum field configurations ranging from  the Abelian constant field (\ref{consfield})  to  a rich chromomagnetic flux tube structure (\ref{magneticsheetsolution1}) permeating through the 3d-space.  In this respect,  it seems important to extend the above consideration of the effective Lagrangian calculation for the general  covariantly constant gauge field configurations.

\section{\it Condensation of chromomagnetic flux tubes}

The proper time integral for the effective  Lagrangian (\ref{yangmillsav}) can be evaluated exactly by using the invariant renormalisation condition \cite{Savvidy:1977,Savvidy:1977as}:
 \be\label{renormcondition1}
 {\partial  \CL \over \partial \CF} \vert_{t = {1\over 2}\ln ({2e^2  \vert \CF \vert \over \mu^4})= \CG=0} =-1, 
 \ee
where $\CF = {1\over 4}G^a_{\mu\nu}G^a_{\mu\nu}$  and $\mu^2$ is the renormalisation scale parameter.   This renormalisation of the Lagrangian for the $SU(N)$ gauge group  gives   (\ref{savvacgg})
\beqa\label{univelagran}
\CL_{YM}^{(1)} = 
-\CF - {11 N  \over 96 \pi^2} g^2 \CF \Big( \ln {2 g^2 \CF \over \mu^4}- 1\Big)~,~~~ \CF =   {\vec{\CH}^2_a - \vec{\CE}^2_a \over 2} >0,~
\CG =  \vec{\CE}_a \vec{\CH}_a =0~.~~
\eeqa
The effective Lagrangian  allows to calculate the quantum-mechanical corrections to the energy-momentum tensor by using the formula derived by Schwinger in \cite{Schwinger:1951nm}:
\beqa\label{tmunu}
T_{\mu\nu} &=& (F_{\mu\lambda} F_{\nu\lambda}-g_{\mu\nu} {1\over 4} 
F^2_{\lambda\rho} ) {\partial \CL \over \partial \CF}- g_{\mu\nu} (\CL -
\CF {\partial \CL \over \partial \CF} - \CG {\partial \CL \over \partial \CG}). 
\eeqa
 Let us first consider the contribution of massless quarks to the effective Lagrangian. In the massless limit the  Heisenberg-Euler  effective Lagrangian in QED has the exact logarithmic dependence as the function of  the invariant $\CF$ \cite{Savvidy:2019grj}:
\beqa\label{QED0}
 \CL_e
&=& -\CF + {   e^2 \CF  \over 24 \pi^2} 
\Big[  \ln ({2 e^2 \CF \over \mu^4})  - 1  \Big] ,~~~~~~~~~ \CF =   {\vec{\CH}^2 - \vec{\CE}^2 \over 2} \geq 0,~~~
\CG =  \vec{\CE} \vec{\CH} =0,
\eeqa
where $\vec{\CH}$ and $\vec{\CE}$ are magnetic and electric fields, and for $T_{\mu\nu}$ one can obtain 
\beqa\label{EnergyMomenQED0}
T_{\mu\nu} &=& T^{Max}_{\mu\nu}\Big[1 - {   e^2   \over 24 \pi^2} 
  \ln {2 e^2 \CF \over \mu^4}   \Big]
+ g_{\mu\nu}    {   e^2   \over 24 \pi^2} \CF  ,~~~~~~~~~~~~\CG=0,
\eeqa
where $T_{\mu\nu}$  contains the space-time metric tensor $g_{\mu\nu}$ and induces  the positive effective cosmological constant.
It follows  from (\ref{QED0}) that the corresponding quark contribution to the effective Lagrangian in the chiral limit  is
\beqa\label{chirallimit}
 \CL_q&=& -\CF + {    N_f  \over 48 \pi^2}  g^2 \CF
\Big[  \ln ({2 g^2 \CF \over \mu^4})  - 1  \Big] ~,~
\eeqa
where $N_f$ is the number of quark flavours.   The energy-momentum tensor $T_{\mu\nu}$ in the pure $SU(N)$ YM theory can be obtained  from (\ref{univelagran})   and (\ref{chirallimit}): 
\beqa\label{energymomentumYM0}
T_{\mu\nu} = T^{YM}_{\mu\nu}\Big[1 +{ b \ g^2 \over 96 \pi^2} 
  \ln {2 g^2 \CF \over \mu^4} \Big]
- g_{\mu\nu}    { b \   g^2 \over 96 \pi^2}  \CF  ,~~~~~~~~~\CG=0, 
\eeqa
where $b = 11N -2N_f$. The vacuum energy density $T_{00} \equiv  \epsilon(\CF)$ has therefore the following form:
\be\label{energyexpYM0}
\epsilon(\CF)=  ~\CF + {b\ g^2\over 96 \pi^2}  \CF \Big( \ln {2 g^2 \CF \over \mu^4}- 1\Big),
\ee
and  has the minimum at the Lorentz and renormalisation group invariant field-strength \cite{Savvidy:1977as}:
\be\label{chomomagneticcondensate0}
\langle  2 g^2 \CF \rangle_{vac}=    \mu^4  \exp{(-{96 \pi^2 \over b\ g^2(\mu) })}= \Lambda^4_{QCD}. 
\ee
The expression for the effective Lagrangian can be obtained also by solving the renormalisation group equation in terms of the effective coupling constant $\bar{g}(g,t)$  \cite{Savvidy:1977as,Matinyan:1976mp}:
\be\label{effectivecouplingel1}
{\partial \CL \over \partial \CF} = - {g^2 \over \bar{g}^2(t)},~~~~~~~~~{d \bar{g} \over dt } = \beta(\bar{g})~,~~~~~~~~t = {1\over 2}\ln(2 g^2 \CF/ \mu^4),
\ee
and allows to calculate different observables of physical interest, including  the quantum energy momentum tensor, vacuum energy density, the magnetic permeability as a function of sourceless gauge fields. The influence of the chromomagnetic condensation on the cosmological evolution were considered in \cite{Savvidy:2021ahq}.     

I would like to thank Jan Ambjorn, Maxim Chernodub, Henryk Arodz and Konstantin Savvidy for stimulating discussions and email communication.

\section{\it Appendix A. Solution of covariantly constant field equation }

As we have seen, the equation (\ref{YMeqcovint1}) reduces to the following system of partial differential equations  (\ref{gen}):
\beqa\label{gen1}
S_{12}= \sin \theta (\partial_1 \theta \partial_2 \phi - \partial_2 \theta \partial_1 \phi ), \nn\\
S_{23}= \sin \theta (\partial_2 \theta \partial_3 \phi - \partial_3 \theta \partial_2 \phi ), \nn\\
S_{13}= \sin \theta (\partial_1 \theta \partial_3 \phi - \partial_3 \theta \partial_1 \phi ),
\eeqa
where $S_{ij}$ are some constants. The linear combination of these equations 
\be\label{phieq}
 S_{12}\partial_3 \phi   + S_{23} \partial_1 \phi  +S_{31} \partial_2 \phi=0
\ee
 vanishes and defines the  angle $\phi$ as an arbitrary function of  the variable 
$Y=  b_1 x +b_2 y + b_3 z $,
$$ 
 \phi(Y)^{'}_Y~ (S_{12} b_3   + S_{23} b_1  +S_{31} b_2)=0,
$$
thus
$
 \phi(Y)  =\phi( b\cdot x ),
$
where $b_{i}, i=1,2,3$ are arbitrary real numbers defining $S_{ij}$ as the solutions of the above equation.  Substituting the above function into the equations (\ref{gen1}) we will get that
$$
S_{12}   + S_{23}   +S_{31} =  - \phi(Y)^{'}_Y~  \Big((b_2-b_3)  \partial_1 \cos  \theta   +(b_3-b_1)  \partial_2 \cos  \theta   +  (b_1-b_2) \partial_3 \cos  \theta  \Big).
$$
Therefore 
$$
\phi(Y)^{'}_Y~ = -\frac{S_{12}   + S_{23}   +S_{31}}{\Big((b_2-b_3)  \partial_1 \cos  \theta   +(b_3-b_1)  \partial_2 \cos  \theta   +  (b_1-b_2) \partial_3 \cos  \theta  \Big)},
$$
and in order to fulfil the equation (\ref{phieq}) it should remain an arbitrary function of a linear combination of the space coordinates. It follows then that the expression in the brackets also should be an arbitrary function of a linear combination of the space coordinates. In that case the angle variable $\theta$ should be a function of any other linear combination of the space coordinates $X= a\cdot x $, so that
$
 \theta(X)=  \theta( a\cdot x ),
$
where $a_{i}, i=0,1,2,3$ are arbitrary real numbers as well.  It follows then that the equations (\ref{gen1})  reduce to the following system of equations:
\be\label{genecovfie1}
S_{ij} = a_i \wedge b_j  \sin\theta(X) ~ \theta(X)^{'}_X   ~ \phi(Y)^{'}_Y , 
\ee
where the derivatives are over the respective arguments.  The solutions with a constant tensor $S_{ij}$  should fulfil the following equation:
\be\label{ansatz71}
\sin\theta(X) ~ \theta(X)^{'}_X   ~ \phi(Y)^{'}_Y =1,
\ee
so that 
$
S_{ij} = a_i \wedge b_j  
$
and the field-strength tensor and the energy density  will have the following form:
\be\label{genenergden1}
G_{ij}= F_{ij} + {1\over g}  a_i \wedge b_j,~~~~~~\epsilon =  {1\over 4 }G^{a}_{ij} G^{a}_{ij}  =   {(g  \vec{H} -   \vec{a} \times \vec{b} )^2 \over 2 g^2}.
\ee
The minus sign in the brackets is when three vectors  $ ( \vec{a}, \vec{b},\vec{H})$ are forming the orthogonal right-oriented frame and the plus sign for the left-oriented frame \cite{Savvidy:2024sv, Savvidy:2024ppd,Savvidy:2024xbe}.
The variables in (\ref{genecovfie1})  are independent, therefore we can choose an arbitrary function  $\theta$ and define the function $\phi$ by integration. Let $\theta(X) $ be an arbitrary function of $X$, then $\phi = Y/\sin \theta(X) \theta(X)^{'}_X$, and we have the following general solution for the colour unit vector (\ref{unitvector}):
\be\label{generasol1}
n^a(\vec{x})= \{\sin \theta(X)  \cos\Big({Y \over \theta(X)^{'} \sin \theta(X)} \Big),~\sin \theta(X)  \sin\Big({Y \over \theta(X)^{'}  \sin \theta(X) }\Big),~ \cos \theta(X)   \}.
\ee 
Notice that the function  $\phi = Y/\sin \theta(X) \theta(X)^{'}_X$ depends on a linear combination of the coordinates and therefore fulfils the equation (\ref{phieq}). The explicit form of the vector potential $A^a_{\mu}$ can be obtained by substituting the  unit colour vector (\ref{generasol1}) into (\ref{choansatz1}).

Alternative  solutions that have magnetic flux structure were obtained  numerically in \cite{Chernodub:2022ywg} and \cite{Kim:2016xdn, Pak:2020obo,Pak:2020izt,Pak:2017skw, Pak:2020fkt,Pak:2025wci,Tokutake:2025qha}. As far as  they also have periodic magnetic flux structure, it will be interesting to compare these solutions with the covariantly constant solutions.

\section{\it Appendix B.  Structure of chromomagnetic flux tubes}

When $H=0$, the polynomial  solution (\ref{magsheet})  is 
\beqa\label{magsheet1} 
A^{a}_{i}(x,y) = {1\over g} \left\{
\begin{array}{llll}   
 \Big({a \sin b y  \over \sqrt{1-(a x)^2} }  , -{a \cos b y  \over \sqrt{1-(a x)^2} }, 0\Big) \\
 & \\
 b \Big(-a x  \sqrt{1-(a x)^2}  \cos b y , -a x  \sqrt{1-(a x)^2} \sin b y ,   1-(a x)^2 \Big)&\\
(0,0,0),~~~~~~~~~~~~~~~~~~~~~~~~~~~~~~~~~~~~~~~~~~~~~~~~~~~~~~~~~~~~~~~~~~~~~~~~ (a  x)^2 <  1, &
\end{array} \right.   
\eeqa 
where $\vec{a}=(a,0,0)$, $\vec{b}=(0,b,0)$  and $A^{a}_{\mu} =0$ when $ (a  x)^2 >1 \nn$.   The nonzero component of the field-strength tensor has the following form:
\be\label{inducedmag}
G^a_{12}(x,y)= -{a b \over g} \Big( \sqrt{1-(a x)^2}  \cos  b y  ,~ \sqrt{1-(a x)^2}   \sin  b y ,~ a  x   \Big),
\ee
and the corresponding energy density $\epsilon = \frac{a^2 b^2}{2 g^2}$ is a constant.
The calculation of the order parameter $A(L)$ (\ref{magflux})  \cite{tHooft:1981bkw, tHooft:1979rtg,tHooft:1980xss} can be divided into two parts. The gauge field $\hat{A}_1= A^2_1 (x,y) \frac{\sigma_2}{2} =  -{a \cos b y  \over \sqrt{1-(a x)^2} } \frac{\sigma_2}{2}  $ on the lines $y=0,\pi/b, 2\pi/b$ is
\be
A^2_1 (x,0) = - {1\over g}  {a   \over \sqrt{1-(a x)^2} }, ~~A^2_1 (x,\pi/b) = {1\over g}  {a   \over \sqrt{1-(a x)^2} }, ~~A^2_1 (x,2\pi/b) = -{1\over g}  {a   \over \sqrt{1-(a x)^2} }, \nn
\ee
while $A^1_1 (x,0) =A^1_1 (x,\pi/b) =A^1_1 (x,2\pi/b) = A^a_3(x,y) =0 $. On the boundary lines $ x=\pm \frac{1}{a}  $ the gauge field is  $A^a_2(\pm \frac{1}{ a},y)=0$.   Let us consider the closed loop $L_1$ surrounding the oriented magnetic flux tube of the square area $ { 2 \pi \over a b}$ in the $(x,y)$ plane of the solution  (\ref{magsheet1}) (see Fig.\ref{fig11}).  The  phase factor over the contour $L_1: y=0, x\in (1/a,-1/a)$;~ $ y=\pi/b,  x\in (-1/a,1/a)$  is
\be
 \oint_{L_1 } \hat{A}_{\mu} d x_{\mu}= - \int_{1/a}^{-1/a}   {   \sigma_2  a d x \over 2 g \sqrt{1-(a x)^2} }+ \int_{-1/a}^{1/a}   {  \sigma_2 a   d x \over 2 g \sqrt{1-(a x)^2} }= \frac{ \pi}{g} \sigma_2 ,
\ee
and the functional $A(L)$ measuring the magnetic flux through the contour  $L_1$  \cite{tHooft:1981bkw, tHooft:1979rtg,tHooft:1980xss} is\footnote{$2 W(L)$ is a character of the SU(2) representations $\chi_j = {\sin(j+1/2) \Phi  \over \sin(\Phi/2) }$ and for $j=1/2$ is $ \chi_{1/2} = 2\cos(\Phi/2)$. }
\be\label{fluxdef12345}
A(L_1 )={1\over 2} Tr P \exp{(i  g \oint_{L_1} \hat{A}_k d x_k)} = {1\over 2} Tr e^{i g \frac{ \pi}{g}  \sigma_2}  =\cos{\Big({1\over 2}  g  \Phi_1\Big)} =-1,
\ee
where $\Phi_1={2\pi \over g}$.   Considering the second contour $L_2: y=\pi/b, x\in (1/a,-1/a)$,~ $ y=2\pi/b,  x\in (-1/a,1/a)$  of the area $ { 2 \pi \over a b}$ we will obtain  the negative phase factor 
\be
 \oint_{ L_2} \hat{A}_{\mu} d x_{\mu}= \int_{1/a}^{-1/a}   { \sigma_2 a  d x \over 2 g \sqrt{1-(a x)^2} }-  \int_{-1/a}^{1/a}   { \sigma_2 a  d x \over 2 g \sqrt{1-(a x)^2} }= - \frac{ \pi}{g}  \sigma_2.
\ee
 The chromomagnetic fluxes have {\it opposite orientations}, and this fact can be illustrated by computing the total flux through  the loop $L_1\cup L_2$ \footnote{ The  gauge-invariant flux defined in (\ref{magflux}), (\ref{fluxdef12345}) is not in general  an additive quantity. }:
\be
 \oint_{L_1 \cup L_2} \hat{A}_{\mu} d x_{\mu}= - \int_{1/a}^{-1/a}   { \sigma_2 a  d x \over 2 g \sqrt{1-(a x)^2} } - \int_{-1/a}^{1/a}   { \sigma_2 a  d x \over 2 g \sqrt{1-(a x)^2} }= 0, 
\ee
so that $A(L_1 \cup L_2) = 1$. Thus we have a  flux cancellation through the  union of two  cells $x \in (-1/a,1/a), y \in (0, 2\pi/b)$ of the area $\frac{4 \pi}{a b}$. The magnetic flux induced by the constant Abelian field $A_1 = - H y$  through the identical area 
$ { 2 \pi \over a b}$ is
\be\label{fluxdef1234}
A(L )={1\over 2} Tr P \exp{(i  g \oint_{L} \hat{A}_k d x_k)} = {1\over 2} Tr e^{- i g H \frac{ 2\pi}{a b}  \frac{\sigma_1}{2}}  =\cos{\Big({ \pi \over a b }  g H \Big)}.
\ee

\section{\it Appendix C. Properties of the orthonormal frames } 

The vectors $(n^a, \partial_{x} n^a, \partial_{y} n^a)$ are orthogonal:
\be
n^a ~\partial_{x} n^a = n^a ~\partial_{y} n^a =
 \partial_{x} n^a ~ \partial_{y} n^a =0,
\ee
and can be normalised:
$
n^a ~n^a  =1,~~~\partial_{x} n^a~\partial_{x} n^a =\frac{a^2}{1 -(ax)^2},~~~
\partial_{y} n^a~\partial_{y} n^a = b^2(1 - (ax)^2) ,~~
$
so that the orthonormal frame $(n^a, e^a_{1}, e^a_{2})$ is
\be
n^a,~~~e^a_{1}  ={1\over a} \sqrt{1 - (ax)^2} ~\partial_{x} n^a,~~~
e^a_{2} = \frac{1}{b \sqrt{1 - (ax)^2} } \partial_{y}  n^a~.
\ee
The useful  matrix elements of the  operator $\hat{n}^{ab}$ in this orthonormal frame are:
\be\label{matrixelements1}
e^a_1\hat{n}^{ab} e_2^{b} =  1,~~~e^a_2\hat{n}^{ab} e_1^{b} =  -1,~~~n^a\hat{n}^{ab} n^b=n^a\hat{n}^{ab} e_1^b=n^a\hat{n}^{ab} e_2^b=e^a_1\hat{n}^{ab} e_1^{b} =e^a_2\hat{n}^{ab} e_2^{b}=0.
\ee
The identical matrix elements are in the frame of the hyperbolic flux tube solution (\ref{orthonormalframe1}). 

\section{\it Appendix D. Absence of negative mode solutions of YM equation}

The negative mode (\ref{NOcontribution}) appears when the Yang-Mills (YM) equation is considered in the {\it linear approximation} (\ref{vectorhamilt1}), (\ref{negmode}). The main question is if the negative mode amplitude $W$ (\ref{negmodampli}) remains as a solution  of {\it  the nonlinear YM equation} in the constant background field $A^{3 }_2\vert^{ext} = H x $.  The negative-mode amplitude  has the following form \cite{Nielsen:1978rm, Ambjorn:1978ff, Ambjorn:1980ms, Arodz:1980gh}:
\be\label{amb}
W_1 = -i W_2 = W = \frac{1}{\sqrt{2}}(w_1 +i w_2),~~~~~~W_3 = W_0 =0,
\ee 
where $w_1(x,y), w_2(x,y) $ are the real and imaginary parts of the charged field $W_{\mu}= \frac{1}{\sqrt{2}}(A^1_{\mu} +i A^2_{\mu})$.  The components  of the gauge field $A^a_{\mu}$ in the subspace of the negative mode $W$ are:
\beqa\label{gaugefieldcom}
&&A_0= \{ 0,0,0\}\nn\\
&&A_1= \{w_1(x,y), w_2(x,y),0\}\nn\\
&&A_2 =\{-w_2(x,y), w_1(x,y),Hx \}\nn\\
&&A_3 =\{ 0,0,0\},
\eeqa
 We are looking for a nontrivial solution of the YM equation for the fields $w_{i}(x,y), i=1,2$. The nonzero component  of the field strength tensor is 
\be\label{sf6}
G^a_{12} =\left\{-g H x w_2 - \frac{\partial{w_1}}{\partial y}  - \frac{\partial{w_2}}{\partial x},~ g H x w_1 + \frac{\partial{w_1}}{\partial x}-\frac{\partial{w_2}}{\partial y}, H-g \left(w_1^2+w_2^2\right)\right\}, 
\ee
while for the energy density we have
\be\label{fs6}
\epsilon = \frac{1}{2}(H-g \left(w_1^2+w_2^2\right)^2 + (g H x w_2 + \frac{\partial{w_1}}{\partial y}  + \frac{\partial{w_2}}{\partial x})^2 +
(g H x w_1 + \frac{\partial{w_1}}{\partial x}-\frac{\partial{w_2}}{\partial y})^2.
\ee
The background gauge condition $\nabla^{ab}_{\mu} (A^{ext})A^b_{\mu} =0$  takes the following form:
\be\label{gaugecond}
g H x w_2 + \frac{\partial{w_1}}{\partial y}  + \frac{\partial{w_2}}{\partial x}=0,~~~g H x w_1 + \frac{\partial{w_1}}{\partial x}-\frac{\partial{w_2}}{\partial y}=0,
\ee
and leads to the following expressions for the field strength tensor and the energy density:
\be
G_{12} =\left\{0,0, H-g \left(w_1^2+w_2^2\right)\right\},~~~~\epsilon = \frac{1}{2}\left(H-g (w_1^2+w_2^2)\right)^2  .
\ee
The gauge condition (\ref{gaugecond}) simplifies when the components of the negative-mode amplitude are represented as 
\be
w_1 = v_1  e^{-\frac{g H x^2}{2} } ,~~~~~ w_2 = v_2 e^{-\frac{g H x^2}{2} } 
\ee
and  takes the Cauchy-Riemann form:
\be\label{Riemman}
\frac{\partial{v_1}}{\partial y}  + \frac{\partial{v_2}}{\partial x}=0,~~~~~~\frac{\partial{v_1}}{\partial x}-\frac{\partial{v_2}}{\partial y}=0.
\ee
Now the components of the gauge field (\ref{gaugefieldcom}) are  
\beqa
&&A_0= \{ 0,0,0\}\nn\\
&&A_1=  \{v_1(x,y)e^{-\frac{g H x^2}{2} } , v_2(x,y)e^{-\frac{g H x^2}{2} },0\}\nn\\
&&A_2 =   \{-v_2(x,y)e^{-\frac{g H x^2}{2} }, v_1(x,y)e^{-\frac{g H x^2}{2} } ,Hx \}\nn\\
&&A_3 =\{ 0,0,0\}.
\eeqa
Substituting this form of the gauge field into the YM equation  $\partial_1 G^a_{12} - g \epsilon^{abc} A^{b}_{1} G^c_{12}=0$ one can get  the following system:
\beqa
&&  +g^2 v_2 \left(v_1^2+v_2^2\right)-e^{g H x^2} \left(g H \left(v_2-x \left(\frac{\partial{v_1}}{\partial y}+\frac{\partial{v_2}}{\partial x}\right)\right)+ \frac{\partial^2{v_1}}{\partial x\partial y}+\frac{\partial^2{v_2}}{\partial x \partial x}\right) =0,\nn\\
&&-g^2 v_1 \left(v_1^2+v_2^2\right)+e^{g H x^2} \left(g H \left( v_1+x \left(\frac{\partial{v_2}}{\partial y} -\frac{\partial{v_1}}{\partial x}\right)\right)+\frac{\partial^2{v_1}}{\partial x\partial x}-\frac{\partial^2{v_2}}{\partial x \partial y}\right)=0,\nn\\
&&  +2 g H x v_1^2 +v_2  \left( 2 g H x  v_2 - \frac{\partial{v_1}}{\partial y}-3 \frac{\partial{v_2}}{\partial x} \right)+ v_1 \left(\frac{\partial{v_2}}{\partial y}-3 \frac{\partial{v_1}}{\partial x} \right) =0 ,
\eeqa
while the equation $\partial_2 G^a_{21} - g \epsilon^{abc} A^{b}_{2} G^c_{ 21}=0$ gives
\beqa
&& g^2 v_1 \left(v_1^2+v_2^2\right)-e^{g H x^2} \left(g H \left(x \left(\frac{\partial{v_2}}{\partial y}-\frac{\partial{v_1}}{\partial x}\right)+v_1\right)+\frac{\partial^2{v_1}}{\partial y\partial y}+\frac{\partial^2{v_2}}{\partial x \partial y}\right) =0,\nn\\
&&  g^2 v_2 \left(v_1^2+v_2^2\right)+e^{g H x^2} \left(g H \left( x \left(\frac{\partial{v_1}}{\partial y}+\frac{\partial{v_2}}{\partial x}\right)- v_2 \right)+\frac{\partial^2{v_1}}{\partial x\partial y} -\frac{\partial^2{v_2}}{\partial y \partial y}\right) =0,\nn\\
&& v_2  \left(3 \frac{\partial{v_2}}{\partial y} -\frac{\partial{v_1}}{\partial x}\right)+v_1 \left(3 \frac{\partial{v_1}}{\partial y}+\frac{\partial{v_2}}{\partial x} \right) =0.
\eeqa
Due to the gauge conditions (\ref{Riemman}) the equations reduce to the following system:
 \beqa\label{fineq}
&&  g  v_2 \left(v_1^2+v_2^2\right)-e^{g H x^2}   H  v_2     =0,~~~~~~~~ g  v_1 \left(v_1^2+v_2^2\right)-e^{g H x^2}     H  v_1 =0,\nn\\
&&  g H x  \left(v_1^2+v_2^2 \right)    -  v_2 \frac{\partial{v_2}}{\partial x}  -     v_1 \frac{\partial{v_1}}{\partial x}   =0,~~~~ v_2 \frac{\partial{v_2}}{\partial y}  +  v_1 \frac{\partial{v_1}}{\partial y}   =0, \nn\\
&&\frac{\partial{v_1}}{\partial y}  + \frac{\partial{v_2}}{\partial x}=0,~~~~~~\frac{\partial{v_1}}{\partial x}-\frac{\partial{v_2}}{\partial y}=0.
\eeqa
The  solution 
\be\label{triv}
v_1 =0, ~~~~~v_2 =0  
\ee
leads to  $w_1 =w_2=0$ and  the negative-mode amplitude (\ref{amb}) vanishes:  $W=0$. The field strength tensor (\ref{sf6}) and the energy density (\ref{fs6}) reduce to a constant field:
\be\label{triv1}
G_{12} =H, ~~~~~~ \epsilon = \frac{H^2}{2}.
\ee
The   expression $g  \left(v_1^2+v_2^2 \right) = H e^{g H x^2}$ solves the first four equations in (\ref{fineq}) while the components $(v_1,v_2)$ still should fulfil the  Cauchy-Riemann equations (\ref{Riemman}).   Representing these components as
$$ 
v_1(x,y) = \rho(x,y) \cos \phi(x,y),~~~~~~~v_2(x,y) = \rho(x,y) \sin \phi(x,y)
$$
we have $\rho(x) = e^{g H x^2/2} \sqrt{H/g} $, and the Cauchy-Riemann equations  take the following form:
$$
g H x \rho \cos \phi -\rho \sin \phi \partial_x \phi - \rho  \cos \phi \partial_y \phi=0, ~~~-\rho \sin \phi \partial_y \phi + g H x \rho \sin \phi + \rho  \cos \phi \partial_x \phi=0.
$$
The linear combination of these equations gives $  \partial_x \phi  =0$, thus $\phi = \phi(y)$ and both equations reduce to $\partial_y \phi =g H x$. The solution $\phi = g H x y + f(x)$ is inconsistent with $  \partial_x \phi  =0$, therefore $g  \left(v_1^2+v_2^2 \right) = H e^{g H x^2}$ should be rejected as a solution for the negative-mode amplitude.    Thus there are no solutions of the YM equation in the subspace (\ref{amb})   and  $W=0$ \cite{Arodz:1980gh}.

This result does not exclude the existence  of the solutions that are in a subspace different from the $W$  subspace (\ref{amb}).  The  alternative ansatz (\ref{choansatzint}) provides a nontrivial solution of the YM equation in the constant background field that has a magnetic flux tube structure.

\bibliographystyle{elsarticle-num}
\bibliography{magnetic_PLB}

\end{document}